\documentclass[trackchanges, twocolumn]{aastex701}
\usepackage{comment}
\usepackage{amsmath} 
\usepackage{bm}
\usepackage{mathrsfs}
\usepackage{subcaption}
\usepackage{afterpage}
\usepackage{placeins}
\usepackage{hyperref}

\usepackage[T1]{fontenc}

\newcommand{\RO}[1]{\textcolor{red}{[{\bf RO}: #1]}}

\begin{document}

\title{Nonuniform Particle Injection into Black Hole Jets by Radiative Magnetic Reconnection}

\author[orcid=0009-0000-1257-5133]{Rin Oikawa}
\affiliation{Astronomical Institute, Graduate School of Science, Tohoku University, Sendai 980-8578, Japan}
\email[show]{oikawa.rin@astr.tohoku.ac.jp} 

\author[orcid=0000-0002-7114-6010]{Kenji Toma} 
\affiliation{Astronomical Institute, Graduate School of Science, Tohoku University, Sendai 980-8578, Japan}
\affiliation{Frontier Research Institute for Interdisciplinary Sciences, Tohoku University, Sendai 980-8578, Japan}
\email{toma@astr.tohoku.ac.jp}

\author[orcid=0000-0003-2579-7266]{Shigeo S. Kimura}
\affiliation{Astronomical Institute, Graduate School of Science, Tohoku University, Sendai 980-8578, Japan}
\affiliation{Frontier Research Institute for Interdisciplinary Sciences, Tohoku University, Sendai 980-8578, Japan}
\email{shigeo@astr.tohoku.ac.jp}

\begin{abstract}
Active galactic nuclei often exhibit highly collimated relativistic plasma outflows launched from the vicinity of their central black holes. One of the key theoretical challenges in understanding black hole jet formation is the origin of the plasma that feeds the jet, which remains poorly understood, particularly in explaining the observed jet emission.
In this study, we focus on electron–positron pair production generated by high-energy photons from non-axisymmetric magnetic reconnection near the black hole, as suggested by recent three-dimensional general-relativistic magnetohydrodynamics simulations. By employing general relativistic ray tracing, we calculate the spatial distribution of the pair production rate in the jet, taking into account photon propagation and collision angles in curved spacetime.
We find that our scenario can supply a sufficient amount of plasma to explain the observed radio emission from the M87 jet. Furthermore, we show that a spinning black hole plays a crucial role in shaping the spatial distribution of the pairs, which in turn affects jet acceleration and the formation of magnetospheric spark gaps responsible for TeV emission.

\end{abstract}


\section{Introduction}\label{Introduction} 

Black holes in accreting systems are often observed as some of the most luminous objects in the Universe. In some cases, highly collimated relativistic jets are launched from the vicinity of the black hole \citep{Beskin10, Blandford+19}. A prominent example is the supermassive black hole at the center of M87, which has a relativistic jet extending over kiloparsec scales \citep{Hada2024}.

The Blandford–Znajek (BZ) process, which extracts rotational energy from a spinning black hole, is widely regarded as one of the leading mechanisms for powering relativistic jets \citep{Blandford&Znajek77, Komissarov04, Toma2025}. However, the origin of the jet plasma remains poorly understood and constitutes a key unresolved problem in black hole astrophysics.

The composition of black hole jets plays a crucial role in determining their radiative properties (see \citealt{Cerruti20} for a review). In particular, the lepton number density directly affects the emission strength of synchrotron radiation and inverse Compton scattering, which are responsible for the observed radio emission from radio-loud AGN jets and gamma-ray emission from blazars \citep{Kino+15,Ghisellini15}. 
The plasma content also has important dynamical implications: The amount of plasma loaded into the jet determines its inertia, thereby setting an upper limit on the achievable bulk Lorentz factor \citep{Toma&Takahara12,Chow+26}.

Since BZ jets are constructed by large-scale ordered magnetic fields threading the horizon, diffusive transport of charged particles from the disk and wind into the magnetized jet region is strongly suppressed \citep{Mehlhaff2026,Chow+26}. It has therefore been widely argued that photon–photon pair production ($\gamma\gamma \rightarrow e^{+}e^{-}$) can inject plasma into the magnetized jet region \citep{Levinson2011L,Moscribradzka11,Kimura+20,Kisaka+20,Gerorge21,Kin+24}.
In particular, pair production associated with magnetic reconnection in the vicinity of the black hole is a promising source of high-energy photons that may account for the observed jet brightness.\footnote{The pair production by high-energy photons from the accretion disk can supply at most $\sim 100$ times the Goldreich–Julian density \citep{Kimura+20}, and spark gaps keep only the Goldreich–Julian density \citep[e.g.][]{Kisaka+20} (see also Section~\ref{Discussion:VHE}).}

Recent three-dimensional general-relativistic magnetohydrodynamic (GRMHD) simulations, together with two-dimensional general-relativistic particle-in-cell (GRPIC) simulations, suggest that an equatorial current sheet naturally forms and magnetic reconnection is triggered near the black hole in the magnetically arrested disk (MAD) state \citep{Ripperda22,Enzo2026,Mehlhaff2026}. \citet{Enzo2026} argued that, in GRPIC simulations of the MAD state, the accretion flow undergoes intermittent magnetic reconnection near the horizon, which regulates the net accumulation of magnetic flux. When reconnection-driven flux diffusion dominates over flux accumulation, large-scale flux eruptions occur, similar to those reported in GRMHD simulations by \citet{Ripperda22}. 
Such reconnection events efficiently accelerate particles and produce high-energy photons via synchrotron radiation. These photons can create $e^{\pm}$ pairs, providing an efficient plasma supply to the BZ jet with densities reaching $\sim 10^8$–$10^9$ times the Goldreich–Julian density \citep{Ripperda22, Kimura&Toma22, Hokobyan+23, Chne+23, Kuze+24}.

However, previous studies of pair injection via magnetic reconnection have not fully incorporated general relativistic effects. In particular, photon trajectories have often been treated without accounting for null geodesics in curved spacetime, and general relativistic Doppler shifts have not been consistently included. Moreover, the effect of photon collision angles has not been examined in detail. General relativistic light bending may increase the frequency of head-on photon collisions, which could substantially enhance the pair-production rate and modify the spatial distribution of plasma injected into the BZ jet region.

In this paper, we construct a model for plasma injection into the BZ jet driven by radiative magnetic reconnection, explicitly incorporating general relativistic effects. We perform general relativistic ray-tracing calculations to follow photon trajectories and collision angles in curved spacetime. Based on this approach, we derive the pair-production rate and its spatial distribution in the jet. We also take into account the anisotropy of synchrotron photons from magnetic reconnection as suggested by local PIC simulations \citep{Cerutti+14,Chernoglazov+23}. In deriving the pair-production rate, we further consider the physical conditions required to sustain the electric current that drives magnetic reconnection.

This paper is organized as follows. In Section~\ref{radiatice reconnection}, we describe the radiative reconnection model including anisotropy, and the formulation of the pair production rate in curved spacetime. In Section~\ref{sec:method}, we present the numerical method used to compute the spatial distribution of the injected plasma based on general relativistic ray tracing.
Section~\ref{sec:result} contains our main findings: Section~\ref{sec:result_applyingM87} presents the pair-production mechanism and the resulting spatial distribution of injected plasma in the BZ jets; Section~\ref{sec:result_implicationEmission} evaluates the pair-production rate in M87* and discusses whether the injected plasma is sufficient to account for the observed jet emission; Section~\ref{Resuly:GR influence} investigates the influence of black hole spacetime on the spatial distribution; and Section~\ref{Result:Anisotropic influence} examines the spatial distribution and total plasma supply when photon anisotropy is taken into account.
In Section~\ref{sec:Discussion}, we discuss the implications of the injected plasma distribution for jet acceleration and magnetospheric dynamics of BZ jets.

Throughout this paper, we use the notation of $Q_x = Q / 10^x$ in cgs units, except for the black hole mass, $M_x=M/(10^{x}M_{\odot})$.

\section{Model} \label{radiatice reconnection}

\begin{figure}[t]
\raggedleft
\includegraphics[width=\columnwidth]{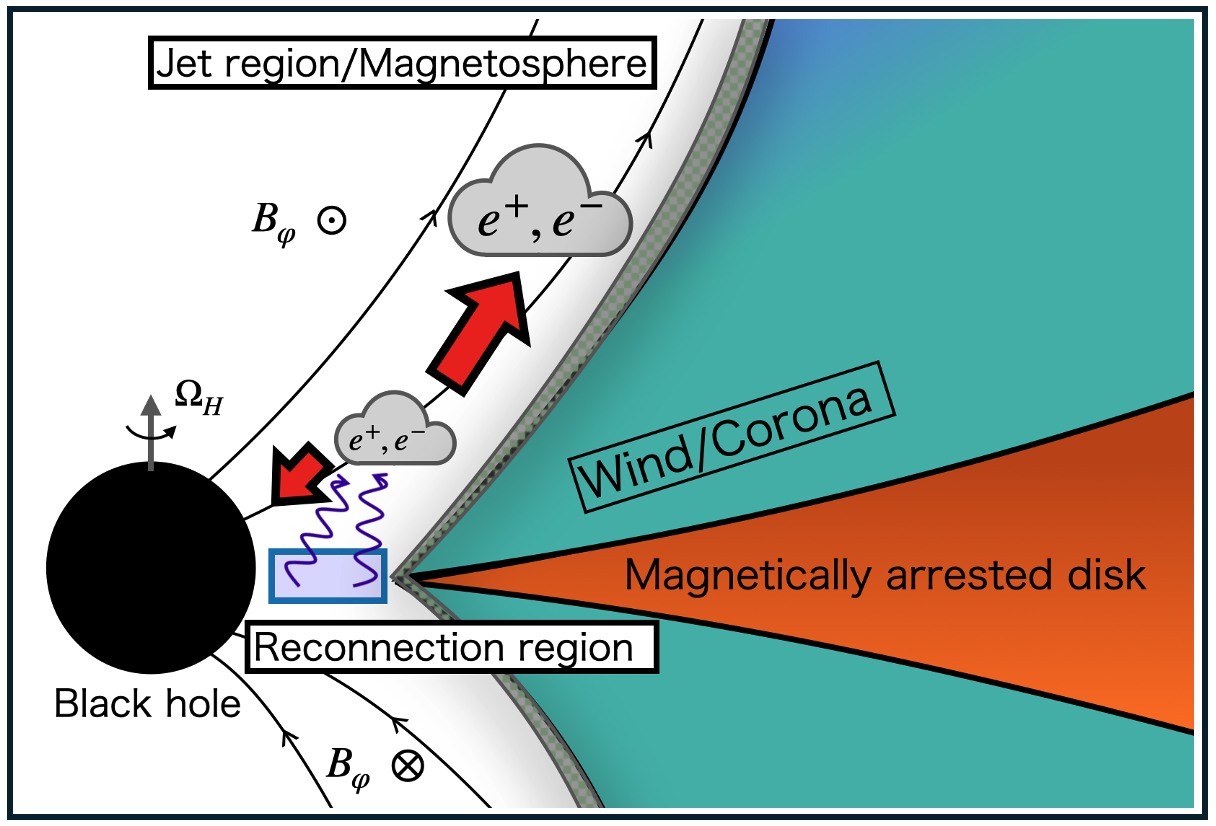}
\caption{Schematic picture of particle injection into the black hole magnetosphere driven by magnetic reconnection. Magnetic reconnection releases magnetic energy and converts it into the energy of nonthermal particles. These nonthermal particles subsequently cool via synchrotron radiation, efficiently producing high-energy photons. The resulting photon field leads to electron–positron pair production. The gray-shaded region represents the transition layer between the magnetosphere and the matter-dominated region (see Section~\ref{Model:self regulate} for details).}
\label{schematic picture}
\end{figure}

Our matter-loading scenario for the BZ jet is shown in Fig.~\ref{schematic picture}. As introduced in Section~\ref{Introduction}, recent 2D global GRPIC simulations by \citet{Enzo2026} have revealed a mini-flare phase, referred to as the reconnection-regulated phase, in which the net magnetic flux near the black hole increases over time while undergoing intermittent magnetic reconnection in the MAD state. Similar mini-flares have also been reported in 3D high-resolution GRMHD simulations \citep{Ripperda22}. These mini-flare releases a large amount of magnetic energy (Section~\ref{Model:EM field}). The reconnection-driven electric field efficiently accelerates particles, converting the released magnetic energy into nonthermal particles \citep{Zenitani+01,Hoshino23, Hoshino+24,Sironi2025}. These particles emit high-energy photons via synchrotron cooling (Section~\ref{Model:photon field}), which then propagate along geodesics in the curved spacetime and undergo pair production ($\gamma\gamma \rightarrow e^{+}e^{-}$) in the black hole magnetosphere. This process steadily loads a substantial amount of plasma into the jet (Section~\ref{Model:pair production}). We assume that the reconnection current sheet has a rectangular shape with a characteristic length scale $l_{\rm rec}=f_l r_g$, where $r_g = GM/c^2$ is the gravitational radius. Here, $G$ is the gravitational constant, $M$ is the black hole mass, and $c$ is the speed of light. We adopt $f_l\approx2$, corresponding to the typical size of the reconnection current sheet in a mini-flare \citep{Ripperda22}. Hereafter, we refer to the current sheet as the `reconnection region'.

In this section, we introduce models for the background electromagnetic field and the photon field produced by radiative magnetic reconnection, and formulate the pair-production rate in general relativity. As the background spacetime, we adopt the Kerr metric, characterized by the black hole mass $M$ and spin parameter $|a|<1$ (explicit expressions for the metric components are given in Appendix~\ref{Appendix:Kerr_spacetime_3+1electrodynamics}).

\subsection{Background electromagnetic field model}\label{Model:EM field}
Magnetic reconnection in the vicinity of a black hole releases a large amount of magnetic energy,
\begin{equation}
    {L}_{\rm rec}\approx{\beta}_{\rm rec}c\frac{{B}^2l_{\rm rec}^2}{4\pi}
    \sim5\times10^{42}\,f_l^2{{\beta}}_{\rm rec,-1}{B}_{3}^2M_9^2\;{\rm erg\,s^{-1}}, \label{eq:released energy}
\end{equation}
where ${\beta}_{\rm rec}\approx0.1$ is the reconnection speed in the kinetic regime \citep[e.g.,][]{Zenitani+01,Lyubarsky+05,Werner+18,Guo+20}, and ${B}$ is the upstream magnetic field strength.
A significant fraction of this energy is converted into nonthermal particles (see Section~\ref{Model:photon field}), depending on the background magnetic field.
In this study, we adopt the analytic model proposed by \citet{Kimura&Toma22} for the background magnetic field in the black hole magnetosphere, in which the poloidal ($B_r$) and toroidal ($B_{\varphi}$) components are roughly $B_r \sim B_{\varphi} \propto r^{-2}$ in the reconnection region considered in this paper (see Appendix~\ref{Appendix:Electromagnetic_field_model}).

\subsection{Local frame of the reconnection layer}\label{Model:local frame}
The energy released by magnetic reconnection is converted into nonthermal
 particles, which subsequently undergo synchrotron cooling and are transformed into radiation energy. To describe the physical quantities associated with these microscopic processes, we introduce a local (i.e., locally Minkowskian) frame in the reconnection region.
 
A spinning black hole induces an electric field 
(see Appendix~\ref{Appendix:Electromagnetic_field_model} and Eq.~\eqref{electric field}). 
However, such a background electric field is typically not included in 
local PIC simulations that investigate magnetic reconnection.
To construct a microscopic model based on PIC simulations, we should consider the local frame 
in which the background electric field vanishes. 
The electric field vanishes locally in the frame moving with the drift velocity given by \citep{Komissarov04,Toma2014}
\begin{equation}
\hat{\bm v}_d = \frac{\hat{\bm D} \times \hat{\bm B}}{\hat{B}^2},
\end{equation}
(see Appendix~\ref{Appendix:orthonormal_basis}), where $\hat{\bm{{D}}}$ and $\hat{\bm{B}}$ denote the electric and magnetic fields measured by the zero-angular-momentum-observer (ZAMO) (see Appendix~\ref{Appendix:Kerr_spacetime_3+1electrodynamics}), and $\hat{\bm v}_d$ is expressed in the ZAMO orthonormal basis.
In this study, we adopt the drift frame as the reference frame in which particle acceleration and radiation processes associated with magnetic reconnection are described. Throughout this paper, quantities expressed in the orthonormal basis of the ZAMO frame are denoted by a hat (e.g., $A^{\hat{\mu}}$ or $\hat{A}$), while those expressed in the orthonormal basis of the drift frame are denoted by a tilde (e.g., $A^{\tilde{\mu}}$ or $\tilde{A}$).

\subsection{Radiative reconnection model}\label{Model:photon field}
In considering radiative magnetic reconnection, the ratio of the following two physical quantities—the magnetization parameter $\sigma$ and the synchrotron burnoff limit $\gamma_{\rm syn}$
are important for determining the strength of radiative cooling \citep{Uzdensky11,Chernoglazov+23,Sironi2025}.
The two quantities are defined as follows:
\begin{equation}
\sigma=\frac{\tilde{B}^2}{4\pi \tilde{\rho} c^2}, \quad
\gamma_{\rm syn}=\sqrt{\frac{6\pi e\tilde{\beta}_{\rm rec}}{\sigma_T \tilde{B}}},
\end{equation}
where $\tilde{\rho}$ is the mass density of the upstream plasma and $\sigma_T$ is the Thomson cross section.
The Lorentz factor $\gamma_{\rm syn}$ corresponds to the maximum particle energy for which the acceleration timescale due to the reconnection electric field $\tilde{t}_{\rm acc} = {\tilde{\gamma} m c}/({e \tilde{\beta}_{\rm rec} \tilde{B}})$ balances the synchrotron cooling timescale $\tilde{t}_{\rm syn} = {\tilde{\gamma} m c^2}/{P_{\rm syn}} = {6\pi m c}/({\sigma_T \tilde{B}^2 \tilde{\gamma}})$.
Here, $m$ is the rest mass of the charged particle and $P_{\rm syn}$ denotes the synchrotron power emitted by a single particle.

When $\gamma_{\rm syn} > \sigma$, radiative cooling can be regarded as negligible, and particles accelerated in the reconnection region can potentially reach Lorentz factors of $\tilde{\gamma} \sim \sigma$. We refer to this regime as the `weak' cooling regime. In contrast, when synchrotron cooling becomes sufficiently strong ($\gamma_{\rm syn} < \sigma$), most particles can be accelerated only up to Lorentz factors of order $\tilde{\gamma} \sim \gamma_{\rm syn}$. We refer to this regime as the `strong' cooling regime.

Recent local 3D PIC simulations by \citet{Chernoglazov+23} have demonstrated that the energy distribution of accelerated particles is significantly modified by secondary acceleration processes. In particular, a fraction of particles can be accelerated up to $\gamma_{\rm syn}$ even in the weak cooling regime (and up to $\sigma$ in the strong cooling regime), resulting in a broken power-law distribution. 
In addition, \citet{Chernoglazov+23} reported the anisotropy of synchrotron radiation produced in magnetic reconnection. Their simulations show that the emission is beamed along the direction of the large-scale reconnection-driven electric field in the weak cooling regime, while it is beamed along the direction of the upstream magnetic field in the strong cooling regime.

Based on the simulations of \citet{Chernoglazov+23}, 
we model the particle energy distribution in radiative magnetic reconnection as follows:
\begin{equation}  \label{particle_distribution}
\
\frac{dN_e}{d\tilde{\gamma}}
\propto
\begin{cases}
\tilde{\gamma}^{-2}
& (\tilde{\gamma}_{\rm dyn}<\tilde{\gamma} < \tilde{\gamma_b}), \\[6pt]
\tilde{\gamma}^{-p_{\mathrm{tail}}}
& (\tilde{\gamma_b} < \tilde{\gamma} < \tilde{\gamma}_{\mathrm{cut}}),
\end{cases}
\end{equation}
{where $\tilde{\gamma}_{\rm dyn}$ is a low-energy cutoff,
$\tilde{\gamma}_b=\min(\sigma,\gamma_{\rm syn})$, and
$\tilde{\gamma}_{\rm cut}=\max(\sigma,\gamma_{\rm syn})$.}
Here $p_{\rm tail}$ is the power-law index of the high-energy tail, and we adopt 
$p_{\rm tail}=2.7$, as suggested by \citet{Chernoglazov+23}. 
Note that, in the weak-cooling regime, the power-law index adopted in Eq.~\eqref{particle_distribution} is taken to be smaller by $1$ than that obtained by \citet{Chernoglazov+23}. {This is because their result describes the particle spectrum in a local region, whereas the particles experience significant synchrotron cooling as they evolve over a dynamical timescale ($\tilde{t}_{\rm syn}\ll\tilde{t}_{\rm dyn}$)}\footnote{{For typical AGN parameters, $\tilde{t}_{\rm syn}\sim10^{-1}\,\tilde{\gamma}_4^{-1}\tilde{B}_{3}^{-2}\,{\rm s}$, while $\tilde{t}_{\rm dyn}=r_g/c\sim10^4\,M_9\,{\rm s}$. Here, the characteristic Lorentz factor is primarily determined by the magnetization in the current sheet ($\tilde{\gamma}\sim\sigma$). For M87*-like parameters, we find $\sigma\sim10^4$ (see Section~\ref{sec:result}).}}, resulting in a spectrum that is steeper by one in the power-law index. The spectrum of strong cooling regime in \citet{Chernoglazov+23} already incorporates the effect of cooling and can be used without modification. {Our model describes the globally cooled particle distribution and adopts a low-energy cutoff of $\tilde{\gamma}_{\rm dyn}\sim1$. Although electrons may accumulate near this cutoff, they make a negligible contribution to the normalization of the distribution adopted in this work. Our treatment is different from PIC simulation without synchrotron radiation \citep[e.g.,][]{French+23}}

\begin{figure}[t]
\raggedleft
\includegraphics[width=\columnwidth]{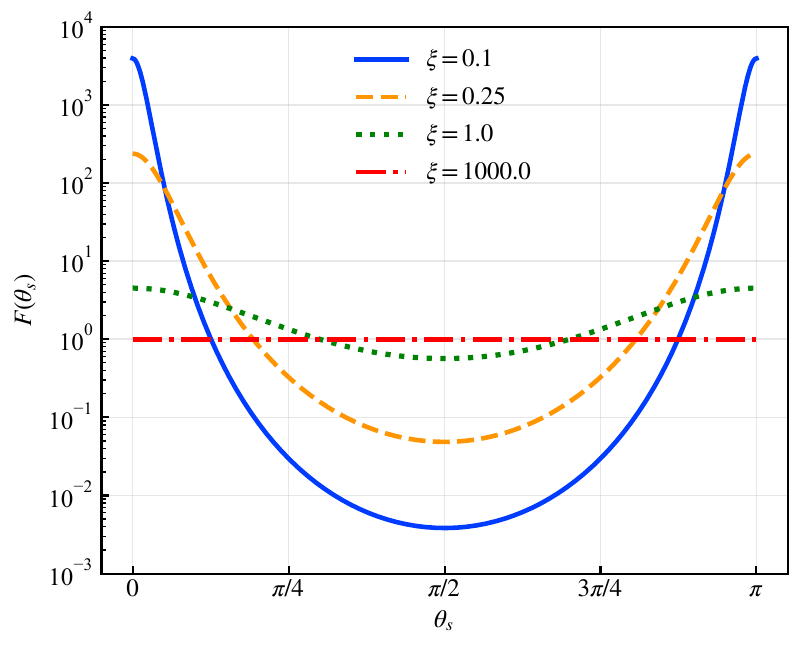}
\caption{Plot of Equation~\eqref{anisotropic function} for the anisotropy parameter $\xi \in (0.1,\,0.25,\,1.0,\,1000)$. Smaller values of $\xi$ correspond to stronger beaming, while larger values indicate that the radiation becomes more isotropic (i.e., $F(\theta_s)\approx1$).}
\label{anisotropic function graph}
\end{figure}

We model the anisotropy of the photon distribution using a functional form\footnote{{The anisotropic function given by Eq.~\eqref{anisotropic function} generally depends on the energy of the emitted photons \citep{Cerutti+14,Chernoglazov+23}. For simplicity, however, we neglect this energy dependence throughout this paper.}} (cf. \citet{Beloborodov+11}:
\begin{equation}
F({\theta_s})\propto \left(\xi^2+\sin^2{\theta_s}\right)^{-3}, 
\label{anisotropic function}
\end{equation}
where $\theta_s$ is the angle between the photon momentum vector ($\tilde{\bm k}$) and the electric field in the drift frame in the weak-cooling regime (or the magnetic field in the strong-cooling regime).
The parameter $\xi$ represents the degree of anisotropy (see Fig.~\ref{anisotropic function graph} for the functional form).

Non-thermal particles rapidly experience synchrotron cooling, efficiently producing a photon field:
\begin{equation}
    G(\tilde{E_\gamma})=\tilde{E_\gamma}\frac{dN_\gamma}{d\tilde{E_\gamma}}\propto
    \begin{cases}
\tilde{E}_{\gamma}^{-\frac{1}{2}}
& (\tilde{E}_{\gamma} < \tilde{E}_{\gamma,b}), \\[6pt]
\tilde{E}_{\gamma}^\frac{1-p_\mathrm{tail}}{2}
& (\tilde{E}_{\gamma,b} < \tilde{E}_{\gamma} < \tilde{E}_{\gamma,\rm cut}),
\end{cases}
\label{eq:Photon spectral}
\end{equation}
where $\tilde{E}_{\gamma} = {h e \tilde{B} \tilde{\gamma}^2}/({2\pi m_e c})$ is the energy of synchrotron photons, and $\tilde{E}_{\gamma,b}$ and $\tilde{E}_{\gamma,\rm cut}$ correspond to emission from electrons with Lorentz factors $\tilde{\gamma}_b$ and $\tilde{\gamma}_{\rm cut}$. We can write the intensity as
\begin{equation}
    \tilde{I}_{\tilde{E}_\gamma}=\tilde{E}_{\gamma}\frac{dN_{\gamma}}{d\tilde{t}d\tilde{\Omega} d\tilde{E}_{\gamma}d\tilde{A}}=C_0F(\theta_s)G(\tilde{E}_\gamma),
\end{equation}
where $d\tilde{A}$ is the surface element, $d\tilde{\Omega} = \sin{\tilde{\theta}} \, d\tilde{\theta} \, d\tilde{\varphi}$ is the solid angle, and $\theta_s$ is related to $\tilde{\theta}$ and $\tilde{\varphi}$ through $\cos\theta_s = \sin{\tilde{\theta}} \cos{\tilde{\varphi}}$.
Here $C_0$ is a normalization factor. {The energy released through magnetic reconnection is almost entirely converted to non-thermal particles \citep{Hoshino23}. In this system, particles undergo efficient synchrotron cooling on global scales ($\tilde{t}_{\rm syn}\ll\tilde{t}_{\rm dyn}$). Furthermore, most particles lose their energy before crossing the horizon because the cooling length is much smaller than the gravitational radius ($c\tilde{t}_{\rm syn}\ll r_g$). As a result, the energy released through magnetic reconnection can be assumed to be converted almost entirely into radiation.} We therefore determine $C_0$ from energy conservation as
\begin{equation}
\tilde{L}_{\rm rec}=2 l_{\rm rec}^2\int_{2\pi} d\tilde{\Omega} \int d\tilde{E}_{\gamma}\;\tilde{I}_{\tilde{E}_\gamma}\cos{\tilde{\theta}},
\end{equation}
where the factor of $2$ accounts for the contributions from both sides of the reconnection current sheet.

We then define the photon distribution function, i.e., the coordinate-invariant six-dimensional phase-space density in the reconnection region, as
\begin{equation}
\mathcal{F}(x_E,k_E) = \frac{c^2}{\tilde{E}_{\gamma}^3} \tilde{I}_{\tilde{E}_{\gamma}}, \label{eq:distribution func}
\end{equation}
where $x_E$ and $k_E$ denote the spacetime position and four-momentum of the photon at the emission point, respectively.

\subsection{$e^{\pm}$ pair production in the BZ jet}\label{Model:pair production}

Photons emitted from the reconnection region propagate through curved spacetime. These trajectories are described by the Hamiltonian formulation of the geodesic \citep{Fuerst04,Pu+16}. The Hamiltonian is given by
\begin{equation}
\mathscr{H}=\frac{1}{2}g^{\mu\nu}k_{\mu}k_{\nu},
\end{equation}
where $g_{\mu\nu}$ is the metric tensor (see Appendix~\ref{Appendix:Kerr_spacetime_3+1electrodynamics} for the explicit form of the Kerr metric), and $k^{\mu}$ is the four-momentum of the photon. The equations of motion which we solve are obtained from Hamilton’s equations,
\begin{equation}
    \frac{dx^{\mu}(\lambda)}{d\lambda} =\frac{\partial\mathscr{H}}{\partial k_{\mu}},\:\:\frac{dk_{\mu}(\lambda)}{d\lambda}=-\frac{\partial\mathscr{H}}{\partial x^{\mu}}, \label{Hamilton eq}
\end{equation}
with $g_{\mu\nu}k^{\mu}k^{\nu}=0$. Here $\lambda$ is an affine parameter along the photon geodesic, and 
$x^{\mu}(\lambda)$ represents the geodesic of the photon.

To evaluate the local $e^{\pm}$ pair production rate, we need the photon distribution function $\mathcal{F}(x(\lambda),k(\lambda))$ at the local point where pair annihilation occurs. We assume that photons emitted from the reconnection region propagate through the magnetosphere along geodesics without undergoing interactions (e.g., with electrons/positrons or other MeV photons) before reaching the annihilation point. This assumption is justified because the optical depths for interactions between photons produced in the reconnection region and the particles and photons in the black hole magnetosphere are much smaller than unity {\citep[][see more details in Appendix~\ref{Appendix:optical depth}]{Kimura&Toma22}}. Their propagation is described by the collisionless Boltzmann equation \citep{Birkl07,Beloborodov11}:
\begin{equation}
    \dfrac{d\mathcal{F}(x(\lambda),k(\lambda))}{d\lambda}=0. \label{eq:Boltzmann_eq}
\end{equation}
{Eq.~\eqref{eq:Boltzmann_eq} shows that $\mathcal{F}(x(\lambda),k(\lambda))$ is conserved along geodesics, and includes the effects of Doppler and gravitational redshifts.} Therefore, the photon distribution function at the local point is equal to that given by Eq.~\eqref{eq:distribution func}, i.e.,
\begin{equation}
    \mathcal{F}(x(\lambda),k(\lambda))=\mathcal{F}(x_E,k_E).
\end{equation}

The local production rate of electron–positron pairs per unit time and unit volume due to photon–photon annihilation is given by \citep{Moscribradzka11,Gerorge21}
\begin{equation}
\begin{split}
\dot{n} &= \frac{1}{\sqrt{-g}} \frac{dN}{d^3 x\, dt} \\
&= \frac{1}{2} \iint \frac{d^3 k}{\sqrt{-g}k^t} 
\frac{d^3 k'}{\sqrt{-g}k'^t} 
\mathcal{F}(x,k)\mathcal{F}(x,k')
\varepsilon_{\rm CM}^2 \sigma_{\gamma\gamma}c,
\end{split}
\label{pair production rate}
\end{equation}
where $g$ is the determinant of $g_{\mu\nu}$. Here $d^3k \equiv dk_1\,dk_2\,dk_3$
denotes the volume element in photon momentum space, where the indices
$1,2,3$ denote the spatial components. $\sigma_{\gamma\gamma}$ is the cross section for $\gamma\gamma \rightarrow e^{+}e^{-}$ \citep{Breit34}:
\begin{equation}
\begin{split}
\frac{\sigma_{\gamma\gamma}}{\sigma_T}
&= \frac{3}{8\varepsilon_{\rm CM}^6}
\left[
(2\varepsilon_{\rm CM}^4 + 2\varepsilon_{\rm CM}^2 -1)
\cosh^{-1}(\varepsilon_{\rm CM})
\right. \\
&\qquad \left.
-
\varepsilon_{\rm CM}(\varepsilon_{\rm CM}^2+1)
\sqrt{\varepsilon_{\rm CM}^2-1}
\right],
\end{split}
\end{equation}
and $\varepsilon_{\rm CM}$ is the photon energy measured
in the center-of-momentum (CM) frame:
\begin{equation}
\varepsilon_{\rm CM}
= -u_{\mu,{\rm CM}}k^{\mu}
= -u_{\mu,{\rm CM}}k'^{\mu}
= \left(\frac{-k_{\mu}k'^{\mu}}{2}\right)^{1/2},
\end{equation}
where $u^{\mu}_{\rm CM}$ is the four-velocity of the CM
frame.
Eq.~\eqref{pair production rate} is invariant under general coordinate
transformations. This is because $\sqrt{-g}\,d^3x\,dt$, $d^3k/(\sqrt{-g}\,k^t)$,
and the distribution function $\mathcal{F}(k)$ are coordinate-invariant, while the photon energy measured by CM and the cross section are scalars. Therefore, the pair production rate $\dot{n}$ is a scalar.

In addition, the local energy--momentum deposition rate due to the pair
production is given by (see e.g., \citet{Moscribradzka11})
\begin{equation}
\begin{split}
\dot{Q}^{\mu} &=
\frac{1}{2}
\iint
\frac{d^3 k}{\sqrt{-g}\,k^t}
\frac{d^3 k'}{\sqrt{-g}\,k'^t}
\,(k^{\mu}+k'^{\mu}) \\
&\qquad \:\:\:\:\:\:\:\:\:\:\:\:\:\:\:\:\:\:\:\times
\mathcal{F}(x,k)\mathcal{F}(x,k')
\varepsilon_{\rm CM}^2
\sigma_{\gamma\gamma}c.
\end{split}
\label{energy_momentum_rate}
\end{equation}

\subsection{A quasi-steady self-regulated state in the reconnection region}\label{Model:self regulate}
At the onset of reconnection, a current sheet forms near the equatorial plane of the accretion disk. The current sheet is initially supplied with plasma from the accretion flow and is therefore likely composed primarily of electrons and ions \citep{Vos2025}. {As the system evolves, the current sheet is fed by plasma from the transition region between the accretion flow ($\sigma_{\rm acc} \sim 1$) and the magnetosphere ($\sigma_{\rm mag} \gg 1$), whose plasma number density gradually decreases with time \citep{Chow+26}. Magnetic reconnection continuously produces high-energy synchrotron photons, which subsequently generate $e^{\pm}$ pairs in the vicinity of the current sheet. As a result, the plasma composition within the current sheet is expected to become increasingly pair dominated over time.}
This pair-loading process self-regulates the upstream magnetization parameter $\sigma$ \citep{Hakobyan2019,Hokobyan+23,Chne+23,Kuze+24}. We therefore examine whether the resulting self-regulated $\sigma$ satisfies the conditions for steady magnetic reconnection.

The condition of steady magnetic reconnection is determined by whether the current sheet can support the required electric current, which is set by Ampère's law. This condition strongly depends on the thickness of the reconnection current sheet. Recent local PIC simulations suggest that the current sheet thickness is roughly of the order of the Larmor radius  in the kinetic regime \citep{Chernoglazov+23,Hokobyan+23,Hakobyan+25}
\begin{equation}
r_{\rm L} = \frac{\langle \tilde{\gamma} \rangle m_e c^2}{e \tilde B},
\end{equation}
where $\langle \tilde\gamma \rangle = {\int d \tilde\gamma\, \tilde\gamma (dN_e/d\tilde\gamma)}/{\int d\tilde\gamma\, (dN_e/d\tilde\gamma)}$ denotes the average Lorentz factor of electrons. This is roughly $\langle \tilde\gamma \rangle \sim \sigma$ for the weak cooling regime and $\langle \tilde\gamma \rangle \sim \gamma_{\rm syn}$ for the strong cooling regime.
From Ampère's law, the current density during magnetic reconnection can be approximated as $\tilde{J} = (c/4\pi)\tilde{\nabla} \times \tilde{B} \approx (c/4\pi)\tilde{B}/r_L$. Equating this with $\tilde{J} \approx \tilde{n}_e e c$, we obtain the number density of charged particles required to carry the electric current in the kinetic regime:
\begin{equation}
\tilde{n}_e\gtrsim \frac{\tilde{B}^2}{4\pi \langle \tilde\gamma \rangle m_e c^2}.
\label{eq:number_density}
\end{equation}

{When $e^\pm$ pairs are produced in the vicinity of the reconnection region,
they are expected to be advected by either the rotation of magnetosphere, their injection velocities, or inflow into the current sheet. The corresponding advection timescale can be approximated
as $f_{\rm adv} r_g/c$, where $f_{\rm adv}$ is a dimensionless parameter. Although $f_{\rm adv}$ is likely to be spatially dependent, we assume it to be constant for simplicity. Therefore, the pair density supplied to the reconnection region is approximately given by $\tilde{n}_e \approx \dot{n}\times f_{\rm adv}{r_g}/{c}$.}
Under this assumption, the condition for the magnetization parameter to self-regulate while simultaneously providing the electric current required to sustain magnetic reconnection can be written as
\begin{equation}
\langle \tilde\gamma \rangle \gtrsim
\frac{\tilde{B}^2}
{4\pi \dot{n} f_{\rm adv}\dfrac{r_g}{c} m_e c^2}
= \sigma.
\label{eq:self_regulate}
\end{equation}
This equation suggests that quasi-steady magnetic reconnection is difficult to sustain in the strong cooling regime.\footnote{\citet{Chernoglazov+23} shows that the current sheet in the strong regime ($\gamma_{\rm syn}=0.2\sigma$) becomes thinner than in the weak cooling regime, although they do not discuss whether the magnetic reconnection can be sustained as a quasi-steady state under condition of even stronger cooling.} If $\sigma$ satisfies Eq.~\eqref{eq:self_regulate}, a quasi-steady reconnection state is expected to be established.

In this study, we determine the value of $\sigma$ near the reconnection region, by which we can calculate the pair creation in the entire magnetosphere.

\subsection{Model parameters}
We summarize our model parameters.
Our model is characterized by five parameters; the black hole mass $M$, the mass accretion rate $\dot{M}$, the spin parameter $a$, the anisotropy parameter $\xi$, and the advection parameter $f_{\rm adv}$.
The magnetic field strength is determined by the black hole mass and the accretion rate under the MAD condition (see Appendix~\ref{Appendix:Electromagnetic_field_model}) ,and is therefore not a free parameter. The spin parameter $a$ governs the structure of the spacetime and the anisotropy parameter $\xi$ determines the strength of the photon anisotropy (see Section~\ref{Model:photon field}). Note that we fix the size parameter of the reconnection region to $f_l = 2$ corresponding to the mini-flare phase, although it may vary between the mini-flare and large-flare phases \citep{Ripperda22}.

We adopt values appropriate for M87*, namely
$M = 6.5 \times 10^9\, M_{\odot}$ and
$\dot{m} = \dot{M}/\dot{M}_{E} = 5.0 \times 10^{-5}$
\citep{EHT2019c,EHT2019f,EHT2021b,Kimura+20,Hada2024},
where
$\dot{M}_{ E} = 1.4 \times 10^{26}\, M_9\;{\rm g\;s^{-1}}$
denotes the Eddington accretion rate.
{We adopt $f_{\rm adv}=1$ throughout this work. We also verified that the pair injection rate is insensitive to the choice of $f_{\rm adv}\gtrsim1$ (see Section~\ref{sec:result_applyingM87}).}
Below, we investigate how $a$ and $\xi$ affect the spatial distribution of injected pairs. The dependence of the pair production rate on $M$ and $\dot{M}$ will be presented in a separate paper.

\section{Method} \label{sec:method}

\begin{figure}[t]
\raggedleft
\includegraphics[width=\columnwidth]{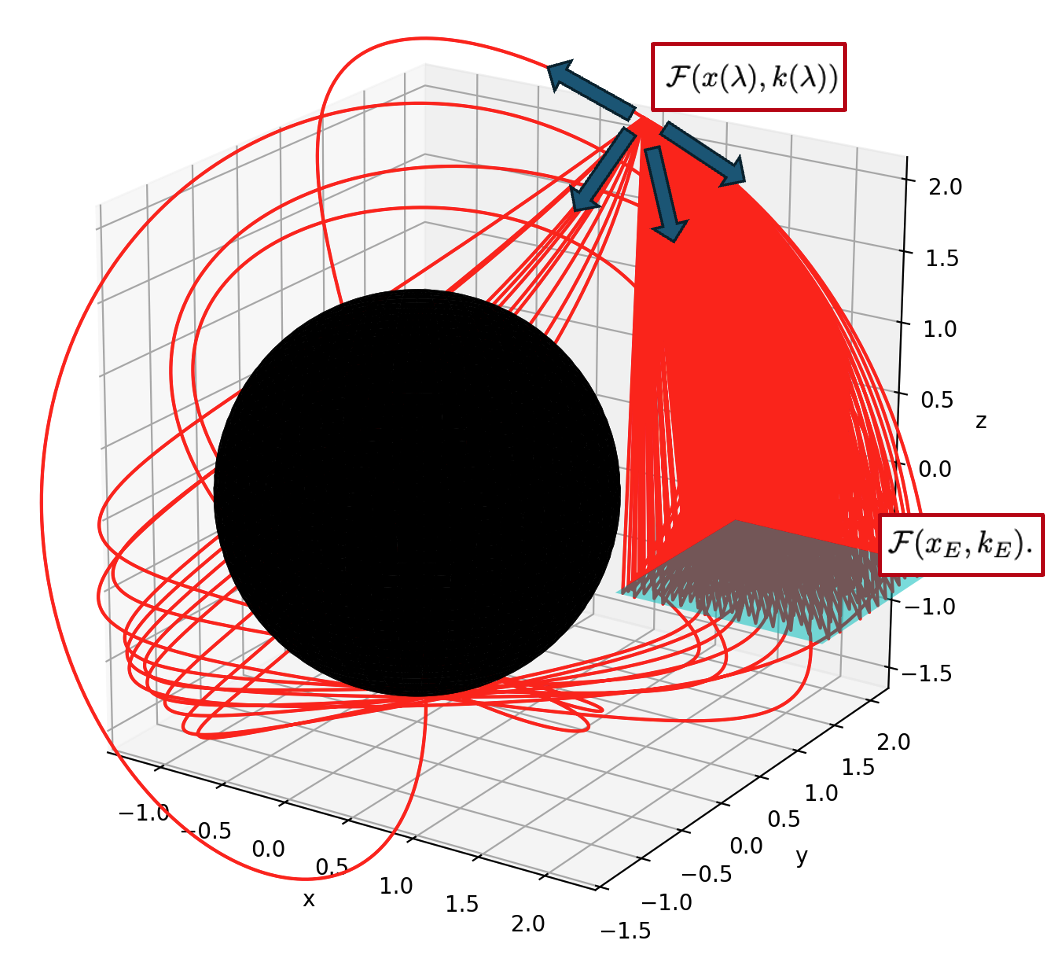}
\caption{Null geodesic ray-tracing in Kerr spacetime. From each point where pair creation occurs, 7,200 photons are isotropically emitted in the ZAMO frame and traced backward in time (red lines). The photon trajectories are followed until they either intersect the reconnection region (cyan region), fall into the black hole horizon, or escape to large radii ($\sim 30\,r_g$).}
\label{Method_picture}
\end{figure}

We compute the pair production rate in each grid cell using a ray-tracing method similar to that of \citet{Birkl07,Beloborodov11}. The spatial grid is constructed in spherical coordinates $(r, \theta, \varphi)$ over the domain $r_h < r \leq r_{\rm max}$, $-\pi/2 \leq \theta \leq \pi/2$, and $0 \leq \varphi \leq 2\pi$, where $r_h = r_g \left( 1 + \sqrt{1 - a^2}\right)$ is the horizon radius. We adopt $r_{\rm max} = 4r_h$ and a grid resolution of $N_r \times N_{\theta} \times N_{\varphi} = 25 \times 21 \times 60$.

The emitting region (i.e., the reconnection region) is assumed to be non-axisymmetric \citep{Ripperda22}. We model it as $1.1r_h \leq x \leq 1.1r_h + l_{\rm rec}$ and $-l_{\rm rec}/2 \leq y \leq l_{\rm rec}/2$. The center of the region is on the $\varphi = 0$ line. We adopt $l_{\rm rec} = 2r_g$, corresponding to a mini-flare \citep{Ripperda22,Enzo2026}.

Photon trajectories are traced backward in time in Boyer--Lindquist coordinates, allowing us to accurately account for their propagation in curved spacetime as well as collision angles (see Fig.\ref{Method_picture}).
Each photon is followed until its geodesic either reaches the reconnection region, falls into the black hole horizon, or escapes to a maximum radius, set to $30\,r_g$. Trajectories reaching the horizon or the outer boundary are assigned $\mathcal{F} = 0$, while those connecting a grid point to the reconnection region are used to evaluate the distribution function $\mathcal{F}=\mathcal{F}(x_E,k_E)$ using Eq.~\eqref{eq:distribution func}.

\subsection{Determination of the Upstream Magnetization}

To compute the pair production rate, the upstream magnetization parameter of the reconnection region must first be determined. As discussed in Section~\ref{Model:self regulate}, the magnetization is self-regulated through pair creation by synchrotron photons.

We determine the upstream magnetization at the center of the reconnection region located at $z = 10^{-4}\,r_g$. Photon trajectories are traced backward in time from this point by isotropically emitting 7,200 photons in the ZAMO frame, and only those intersecting the reconnection region are retained.

For a trial value of $\sigma_{\rm tr}$, we compute the pair production rate $\dot{n}$ and the resulting upstream magnetization parameter $\sigma$ after pair creation. We iteratively update $\sigma_{\rm tr}$ until the following self-regulated condition is satisfied:
\begin{equation}
\left|\frac{\sigma}{\sigma_{\rm tr}} - 1\right| < 0.01.
\label{method:itretation for magnetization parameter}
\end{equation}
Once an upstream magnetization parameter $\sigma$ is obtained, we evaluate whether it satisfies the condition for steady magnetic reconnection (see Eq.~\eqref{eq:self_regulate}):
\begin{equation}
\frac{\langle \tilde\gamma \rangle}{\sigma} \gtrsim 1,
\label{method:itretation}
\end{equation}
where $\langle \tilde\gamma \rangle$ is evaluated at the converged magnetization parameter $\sigma$. This procedure allows us to verify the self-regulated $\sigma$ and its consistency with the condition for steady magnetic reconnection.

\begin{figure*}[t]
\centering
\includegraphics[width=\textwidth]{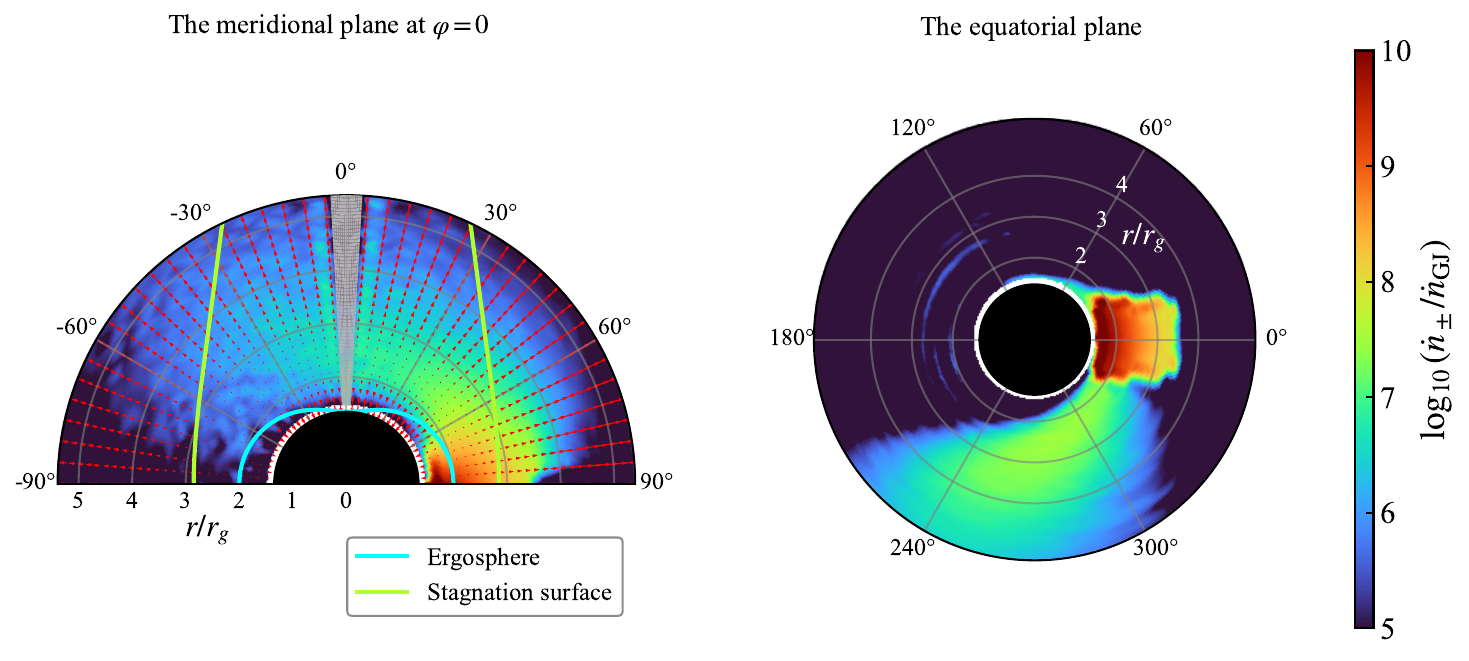}
\caption{Spatial distribution of injected $e^{\pm}$ pairs in the black hole magnetosphere via radiative magnetic reconnection for M87*, with $a = 0.9375$ and $\xi = 1000$. The left panel shows the spatial distribution on the meridional plane at $\varphi = 0$, and the right panel shows the distribution on the equatorial plane. Photon geodesics become singular along the black hole rotation axis ($\theta=0$); accordingly, this region is masked in gray. The colormap shows the multiplicity, defined as the ratio of the pair injection rate, $\dot{n}_{\pm}$, to $\dot{n}_{\rm GJ}=n_{\rm GJ}({c}/{r_g})$,
where $n_{\rm GJ}\approx aB/(8\pi e r_h)$ is the Goldreich--Julian density evaluated at the center of the reconnection region. The light blue line indicates the ergosphere, while the light green line represents the stagnation surface, which separates the inflow and outflow regions for a cold plasma. The red arrows show the projection of the radial velocity $v^{\hat{r}}/c=\dot{Q}^{\hat{r}}/\dot{Q}^{\hat{t}}$. We confirm that the upstream magnetization parameter of the reconnection region is self-regulated by pair creation, $\sigma = 5.5 \times 10^{4}$ (for $f_{\rm adv}=1$).}
\label{results for M87}
\end{figure*}

\subsection{Pair Injection on the Spatial Grid}

Using the obtained magnetization, we compute the pair production rate across the spatial grid. From each grid point, 7,200 photons are isotropically emitted in the ZAMO frame (see Fig.~\ref{Method_picture}), and their trajectories are traced backward in time. Applying this procedure to all spatial grid points yields the spatial distributions of the $e^{\pm}$ pair production rate $\dot{n}$ and the four-momentum deposition rate $\dot{Q}^{{\mu}}$.

\begin{table}[t]
\vspace{10pt}
    \centering
    \begin{tabular}{lc}
        \hline\hline
        Half spatial resolution & Half geodesics \\
        \hline
        $\left|\Delta \dot{N}/\dot{N}\right| = 0.03$
        & $\left|\Delta \dot{N}/\dot{N}\right| = 4 \times 10^{-3}$ \\
        $\left|\Delta L_{\pm}/L_{\pm}\right| = 0.07$
        & $\left|\Delta L_{\pm}/L_{\pm}\right| = 5 \times 10^{-3}$ \\
        \hline\hline
    \end{tabular}
    \caption{Relative errors in $\dot{N}$ and $L_{\pm}$ for half the grid resolution (left column) and half the number of geodesics (right column), compared with the fiducial run.}
    \label{tab:code_accuracy}
\end{table}

\subsection{Code accuracy}
The accuracy of the local values of $\dot{n}$ and $\dot{Q}^\mu$ depends on the number of geodesics launched from each grid cell. A comparison between the fiducial calculation and a reduced calculation with half the number of geodesics shows that the errors of the volume-integrated quantities $\dot{N}$ and $L_\pm$ (see Eqs.~\eqref{eq:pair production rate} and \eqref{eq:energy production rate}) are below a few per cent (see Table~\ref{tab:code_accuracy}).

The accuracy of the volume-integrated quantities $\dot{N}$ and $L_{\pm}$ 
also depends on the grid resolution. A comparison between the fiducial calculation and a run with half the number of grid points shows the differences of a few per cent in $\dot{N}$ and $L_{\pm}$ (Table~\ref{tab:code_accuracy}).

\section{Result}\label{sec:result}

\subsection{Matter loading into the BZ jet}\label{sec:result_applyingM87}

{Here, we present the case of M87* assuming a high spin parameter $a = 0.9375$, and an isotropic photon distribution with $\xi = 1000$ (results for different spin parameters and photon anisotropies are presented in Sections~\ref{Resuly:GR influence} and \ref{Result:Anisotropic influence}, respectively).} {Fig.~\ref{results for M87} shows the spatial distribution of the pair injection rate, $\dot{n}_{\pm}/\dot{n}_{\rm GJ}$, into the magnetosphere produced by radiative magnetic reconnection. Here, $\dot{n}_{\rm GJ}=n_{\rm GJ}({c}/{r_g})$,
where $n_{\rm GJ}\approx aB/(8\pi e r_h)$ is the Goldreich--Julian density evaluated at the center of the reconnection region. 
Note that $e \dot{n}_{\rm GJ}$ is the same order of magnitude as the magnetospheric current density
\citep{Ripperda22, Hokobyan+23}.} We find that the upstream magnetization parameter of the reconnection region is self-regulated to $\sigma = 5.6 \times 10^4$ for $f_{\rm adv}=1$, 
which is much smaller than $\gamma_{\rm syn}=5.1\times10^6$ (i.e., the weak cooling regime). 
This $\sigma$ value is roughly consistent with those estimated without general relativistic effects \citep{Chne+23, Kuze+24}.

The spatial distribution shows that the injected plasma is concentrated near the reconnection region (i.e., the jet boundary), where the photon flux peaks, reaching pair production rate of $\dot{n}_{\pm} \approx 10^9 \dot{n}_{\rm GJ}$. This result is consistent with the limb-brightened feature of the jet, where the emission is enhanced near its edges \citep{Takahashi+18,Lu+2023,Kim+25}. In addition, a substantial number of pairs with $\dot{n}_{\pm} \approx 10^6 \dot{n}_{\rm GJ}$ are supplied to the jet spine near the black hole rotation axis. On the equatorial plane, the spatial distribution exhibits a non-axisymmetric tail produced by frame-dragging effects associated with the rotating black hole. This feature reflects the locations where photons emitted in the prograde and retrograde directions collide most frequently.

{The existence of a nonthermal power-law tail
($\sigma < \tilde{\gamma} < \gamma_{\rm syn}$; see Eq.~\eqref{particle_distribution})
is essential for sustaining matter loading into the jet. Electrons with Lorentz
factors of order $\sigma$ emit synchrotron photons with a characteristic energy
\begin{equation}
\tilde{E}_{\sigma}=\frac{he\tilde{B}}{2\pi m_e c}\sigma^2
=\tilde{E}_{\rm syn}\left(\frac{\sigma}{\gamma_{\rm syn}}\right)^2
\approx 2\,{\rm keV},
\end{equation}
where $\tilde{E}_{\rm syn}=16\,\tilde{\beta}_{\rm rec,-1}\,{\rm MeV}$ is the
synchrotron burnoff-limit photon energy. $E_{\sigma}$ is smaller than the minimum
photon energy required for pair creation with synchrotron burnoff-limit photons,
\begin{equation}
\tilde{E}_{\rm th}\approx\frac{(m_e c^2)^2}{\tilde{E}_{\rm syn}}
\approx 16\,\tilde{\beta}^{-1}_{\rm rec,-1}\,{\rm keV}
>\tilde{E}_{\sigma}.
\end{equation}
Therefore, synchrotron photons emitted by electrons with
$\tilde{\gamma}\sim\sigma$ do not contribute significantly to pair creation. The pair production is driven by synchrotron photons emitted by the
high-energy electrons in the power-law tail, characterized by the index
$p_{\rm tail}$. 
This is the reason why the pair injection rate $\dot{n}$ is insensitive to $f_{\rm adv} \gtrsim 1$. The magnetization parameter $\sigma$ scales as $f_{\rm adv}^{-1}$ (see Eq.~\ref{eq:self_regulate}) and then $\tilde{E}_{\sigma} \propto f_{\rm adv}^{-2}$, but this variation hardly affects the normalization factor $C_0$ or the pair injection rate for $p_{\rm tail}=2.7$.}

\begin{figure}[t]
\vspace{15pt}
\centering
\includegraphics[width=\columnwidth]{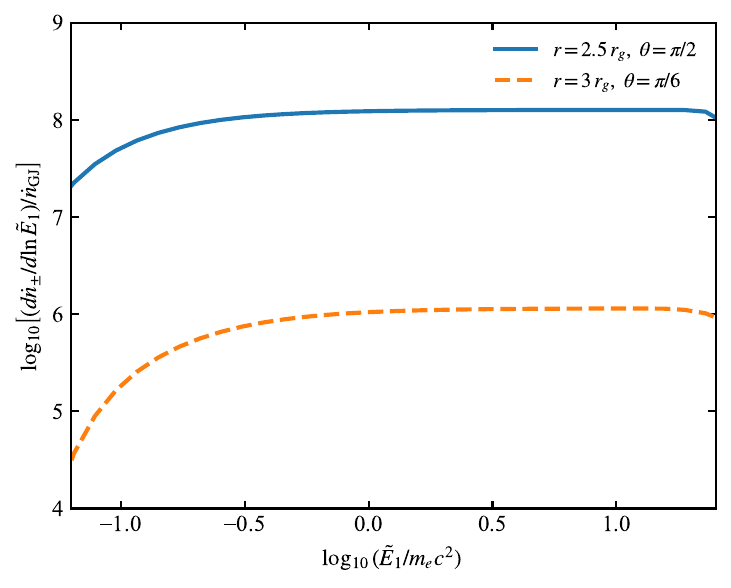}
\caption{
{This figure shows $d\dot{n}_{\pm}/d \ln\tilde{E}_1$, which quantifies the contribution of photons with energy $\tilde{E}_1$ to the pair-production rate per logarithmic energy interval in M87*, with $a=0.9375$ and $\xi=1000$. The blue solid line shows the spectrum at the center of the reconnection region, and the orange dashed line shows the spectrum at $(r,\theta,\varphi)=(3r_g,\pi/6,0)$.}
}
\label{injection spectral}
\end{figure}

{Fig.~\ref{injection spectral} shows $d\dot{{n}}_{\pm}/d \ln\tilde{E}_1$, which quantifies the contribution of photons with energy $\tilde{E}_1$ to the pair production. For a photon of energy $\tilde{E}_1$, the dominant contribution
comes from photons of energy $\tilde{E}_2$ satisfying the threshold condition
$\tilde{E}_1\tilde{E}_2 \sim (m_e c^2)^2$. As shown in Fig.~\ref{injection spectral}, when the photon spectrum follows a power law, each logarithmic energy interval contributes approximately equally to the pair-production rate over the range $\tilde{E}_{\rm th}<\tilde{E}_1<\tilde{E}_{\rm syn}$ \citep[cf. Appendix D of][]{Kimura&Toma22}.}

\subsection{Implications for jet emission and energetics}\label{sec:result_implicationEmission}

\begin{figure*}[t]
\centering
\includegraphics[width=\textwidth]{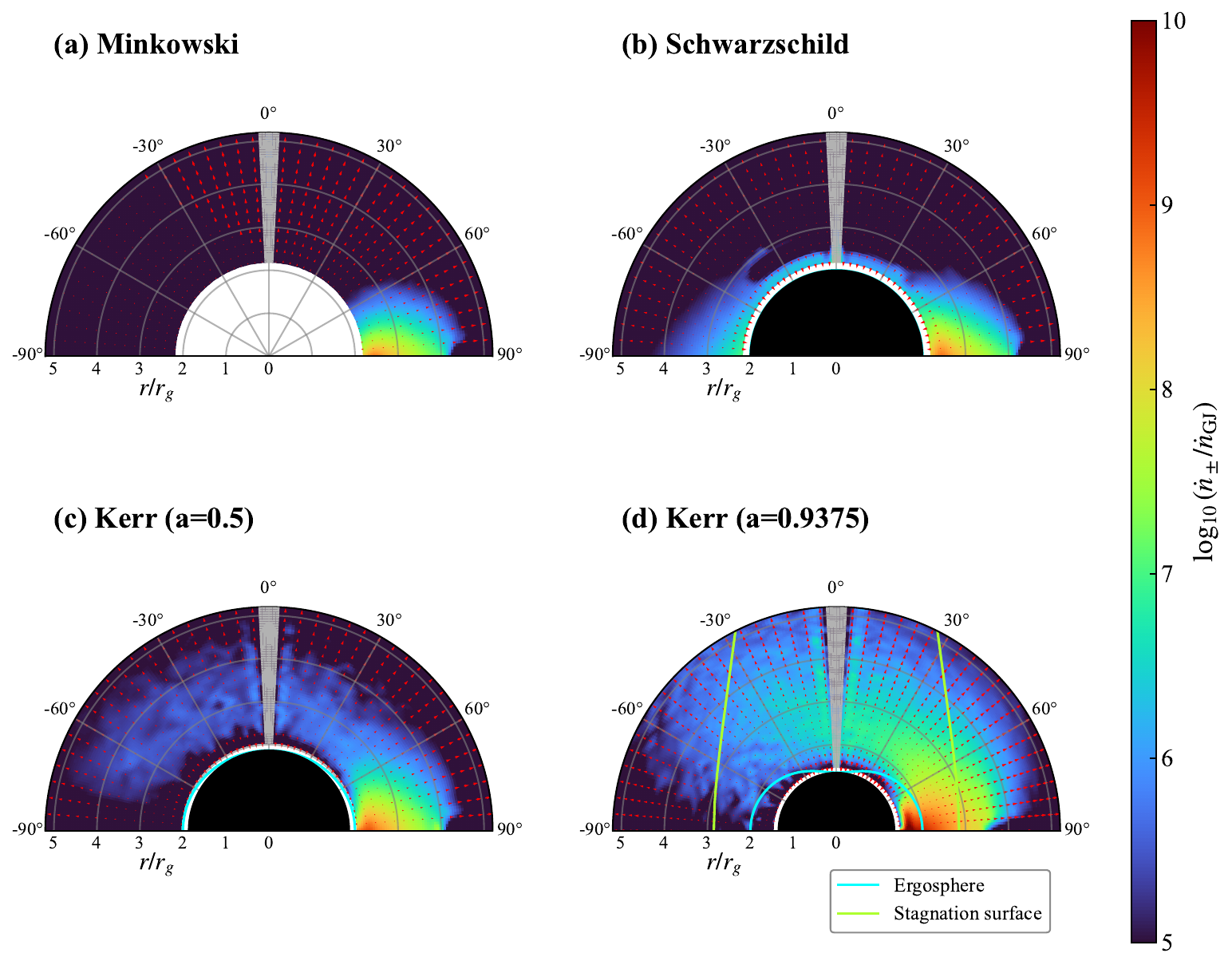}
\caption{Comparison of the spatial distribution of pair injection into the magnetosphere on the meridional plane at $\varphi = 0$ for different spacetime geometries (the equatorial-plane distribution is shown in Appendix~\ref{Appendix:Spartial distribution in equator}, Fig.~\ref{GR2}). Panels (a)–(d) show the density distributions for Minkowski, Schwarzschild, Kerr metric with $a = 0.5$, and $a = 0.9375$, respectively. The Goldreich–Julian density is evaluated using the Kerr black hole with $a = 0.9375$. {In the $a=0.5$ case, the stagnation surface is located outside the current sheet {($r_{\rm stag}\approx6r_g$ on the equatorial plane)}, resulting in most of the high-density pair plasma being being injected within the stagnation surface.
}}
\label{GR1}
\end{figure*}

We examine whether the plasma supplied at the jet base in this scenario can account for the observed brightness of the radio jet. For this estimate, we consider only the pair production rate of plasma supplied outside the stagnation surface ($r > r_{\rm stag}$). The stagnation surface corresponds to the separation point between inflow and outflow under the steady cold-MHD approximation; therefore, plasma injected at $r > r_{\rm stag}$ is generally expected to be carried outward and contribute to the jet. {Although some pairs injected outside the stagnation surface may flow back toward the current sheet along the reconnection inflow as found by \citet{Vos2025}, the injected pairs in our model typically move with velocities comparable to or exceeding the reconnection inflow velocity (see the bottom panel of Fig.~\ref{GR_theta_distribution}).} We therefore expect that the majority of the pairs injected outside the stagnation surface are ejected as the jet.

The corresponding injection rate to the jet is
\begin{equation} \label{eq:pair production rate}
\dot{N} = \int_{r > r_{\rm stag}} \dot{n} \sqrt{-g}\; d^3x \approx 1.0\times10^{43}\ {\rm s^{-1}}.
\end{equation}
This estimate is conservative, as it neglects pairs that may be produced with outward momentum and thus contribute to the outflow even if they are created inside the stagnation surface (see Section~\ref{Discussion:SSA}).

The synchrotron emission power from the jet can be written as
\begin{equation} 
L_{\rm syn} \approx \dot{N}\, T_{\rm dur}\, P_{\rm syn},
\end{equation}
where $T_{\rm dur} \approx r_g/(\beta_{\rm rec} c) \simeq 5.0 \times 10^4\, M_9 \beta_{\rm rec,-1}^{-1}\ {\rm s}$ is the reconnection duration time \citep{Sironi2025}, and $P_{\rm syn} = B_{\rm dis} \sigma_T \nu_{\rm syn} m_e c^2/(3e)$ is the synchrotron power emitted by a single electron. Here, $\nu_{\rm syn}$ is the synchrotron photon frequency, and $B_{\rm dis}$ denotes the magnetic field strength at the dissipation radius $r_{\rm dis}$. Assuming that the toroidal magnetic field dominates in the distant jet region, it can be approximated as $B_{\rm dis} \approx B_{\rm mad} (r_{\rm dis} \sin\theta_j / r_h)^{-1}$, where $B_{\rm mad}$ is the magnetic field strength at the horizon (see Appendix~\ref{Appendix:Electromagnetic_field_model}) and $\theta_j \approx 0.1\,\mathrm{rad}$ is the jet opening angle.

In our scenario, the pairs injected at the jet base produce synchrotron emission at $\nu_{\rm syn} = 43\,\mathrm{GHz}$ with a luminosity of $L_{\rm syn,43\,GHz} \approx 10^{38}\,\mathrm{erg\ s^{-1}}$ at a dissipation radius of $r_{\rm dis} = 10^3 r_g$ in M87*, which is consistent with the previous analytical estimate of \citet{Kimura&Toma22}. VLBI observations at 43 GHz detect radio emission from the jet at $r_{\rm dis} \sim 10^2$--$10^3 r_g$ with luminosities of roughly $10^{37}$--$10^{38}\,\mathrm{erg\ s^{-1}}$ \citep{Walker+18}.  This indicates that our scenario can supply sufficient matter loading to account for the observed radio jet.

Finally, we compare the energy injection rate with the BZ power. The energy injection rate into the magnetosphere is \citep{Beloborodov11}
\begin{equation} \label{eq:energy production rate}
L_{\pm} = -\int_{r > r_h} \left( g_{tt} \dot{Q}^{t} + g_{t\varphi} \dot{Q}^{\varphi} \right) \sqrt{-g}\; d^3x,
\end{equation}
where we obtain \( L_{\pm} \approx 3.2 \times 10^{38}\,\mathrm{erg\,s^{-1}} \) in our numerical calculations.
The BZ power can be expressed as
\begin{equation}
L_{\rm BZ} = \frac{\kappa}{4\pi} \left(\frac{\Omega_H r_g}{c} \right)^2 \Phi_{\rm BH}^2 f(\Omega_H)\dot{M}c^2,
\end{equation}
where $f(\Omega_H)=1+1.38(\Omega_H r_g/c)^2-9.2(\Omega_H r_g/c)^4$, $\Omega_H=ac/(2r_h)$ is the angular velocity of the horizon, $\kappa$ is a magnetic geometry factor, and $\Phi_{\rm BH}$ is the dimensionless magnetic flux threading the black hole (see \citet{Tchekhovskoy2010, Tchekhovskoy11} for details). Adopting $\kappa = 0.053$ and $\Phi_{\rm BH} = 50$, corresponding to a split-monopole configuration and the MAD state \citep{Tchekhovskoy11, Narayan12}, we obtain ${L_{\rm BZ}}/{L_{\pm}} \approx 10^5 \gg 1$ for M87*.
This result indicates that the electromagnetic energy strongly dominates over the particle energy; therefore, pair injection driven by radiative reconnection is unlikely to significantly perturb or disrupt the current structure of the magnetosphere. This is also related with the fact that the resultant magnetization parameter $\sigma \gg 1$.

\subsection{Influence of general relativistic effects}\label{Resuly:GR influence}

We next investigate how the spacetime of a black hole influences the spatial distribution of injected pairs. Even though a non-spinning black hole cannot drive a BZ jet \citep{Enzo2026}, we consider not only Kerr metric with spin parameters $a = 0.5$ and $0.9375$, but also a Minkowski and Schwarzschild metric.

Fig.~\ref{GR1} shows the spatial distribution of injected pairs on the meridional plane at $\varphi = 0$ for each spacetime metric. The magnetization in the reconnection region, self-regulated by pair creation, is summarized in Table~\ref{Comparison with each spacetime}, and we confirm that all cases lie in the weak cooling regime. {We note that the Schwarzschild and Minkowski cases are included primarily as idealized reference models to isolate the effects of spacetime geometry and black hole spin. Therefore, the corresponding magnetizations should not be interpreted as realistic predictions for non-spinning systems.}
 
\begin{figure}[t]
\vspace{15pt}
\centering
\includegraphics[width=\columnwidth]{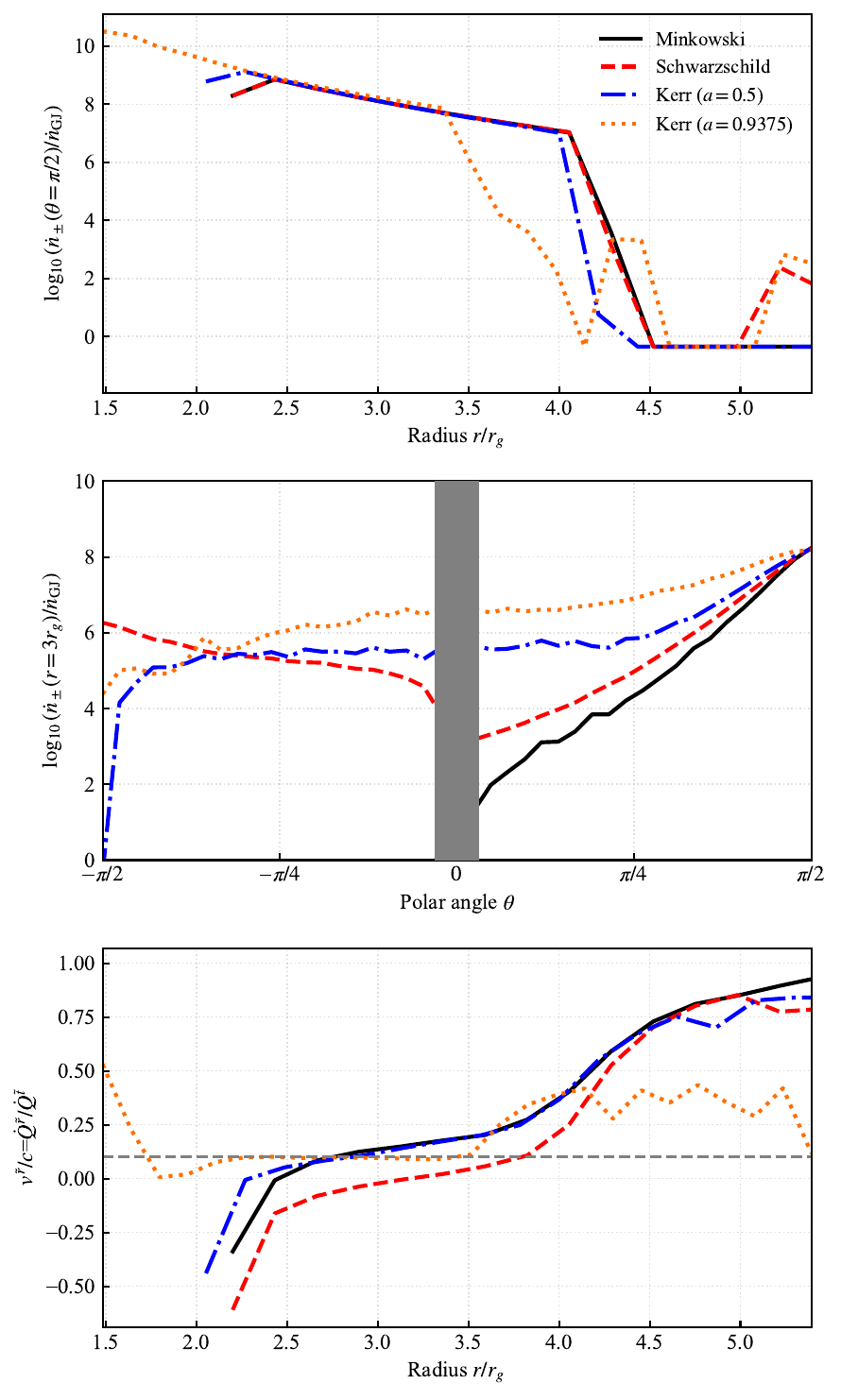}
\caption{{Pair production rate and velocity distribution in the cases of Fig.~\ref{GR1}. The top panel shows the radial ($r$) distribution of the pair production rate on the equatorial plane, the middle panel shows its polar-angle ($\theta$) distribution at $r=3r_g$ and the bottom panel shows the radial velocity ($\dot{Q}^{\tilde{r}}/\dot{Q}^{\tilde{t}}$) of the injected pairs measured in the drift frame at $\theta=\pi/2$. The gray dashed line in the bottom panel denotes the reconnection inflow velocity ($\tilde{\beta}_{\rm rec}=0.1$).}}
\label{GR_theta_distribution}
\end{figure}

 We find that the pair injection near the reconnection region is largely insensitive to the spacetime metric, yielding nearly identical values across the different models (see the top panel of Fig.~\ref{GR_theta_distribution}). In contrast, as clearly shown in Fig.~\ref{GR1}, the spatial distribution of the injected pairs away from the reconnection region exhibits a strong dependence on the spacetime geometry.

\begin{table}[t]
    \centering
    \vspace{20pt}
    \begin{tabular}{lccc}
        \hline\hline
        Metric
        & ${\sigma}$
        & $\gamma_{\rm syn}$
        & $\dot{N}_{r>r_{h}}\;\left(\,\rm s^{-1}\right)$ \\
        \hline
        Minkowski & $1.7\times10^5$ & $6.7\times10^6$ & $3.6\times 10^{42}$   \\
        Schwarzschild & $1.7\times10^{5}$ & $6.7\times10^6$ & $6.2\times 10^{42}$   \\
        Kerr ($a=0.5$) & $1.4\times10^5$ & $6.4\times10^6$ & $1.2\times 10^{43}$   \\
        Kerr ($a=0.9375$)& $5.6\times10^4$ & $5.1\times10^6$ & $6.0\times 10^{43}$  \\
        \hline\hline
    \end{tabular}
    \caption{Comparison of the self-regulated upstream magnetization parameter for $f_{\rm adv}=1$ and the synchrotron burn-off limit of the reconnection region, and the pair production rate outside the horizon for different spacetime metrics. In all cases, the self-regulated magnetization satisfies $\sigma < \gamma_{\rm syn}$. {Note that adopting $f_{\rm adv}>1$ simply rescales $\sigma$ by a factor of $f_{\rm adv}^{-1}$, while the pair production rate does not change.}  {In the $a=0.5$ case, the pair production rate outside the event horizon is sufficient to account for the radio emission. Nevertheless, the majority of the injected pairs are produced within the stagnation surface.}}
    \label{Comparison with each spacetime}
\end{table}

\begin{figure*}[t]
\centering
\includegraphics[width=\textwidth]{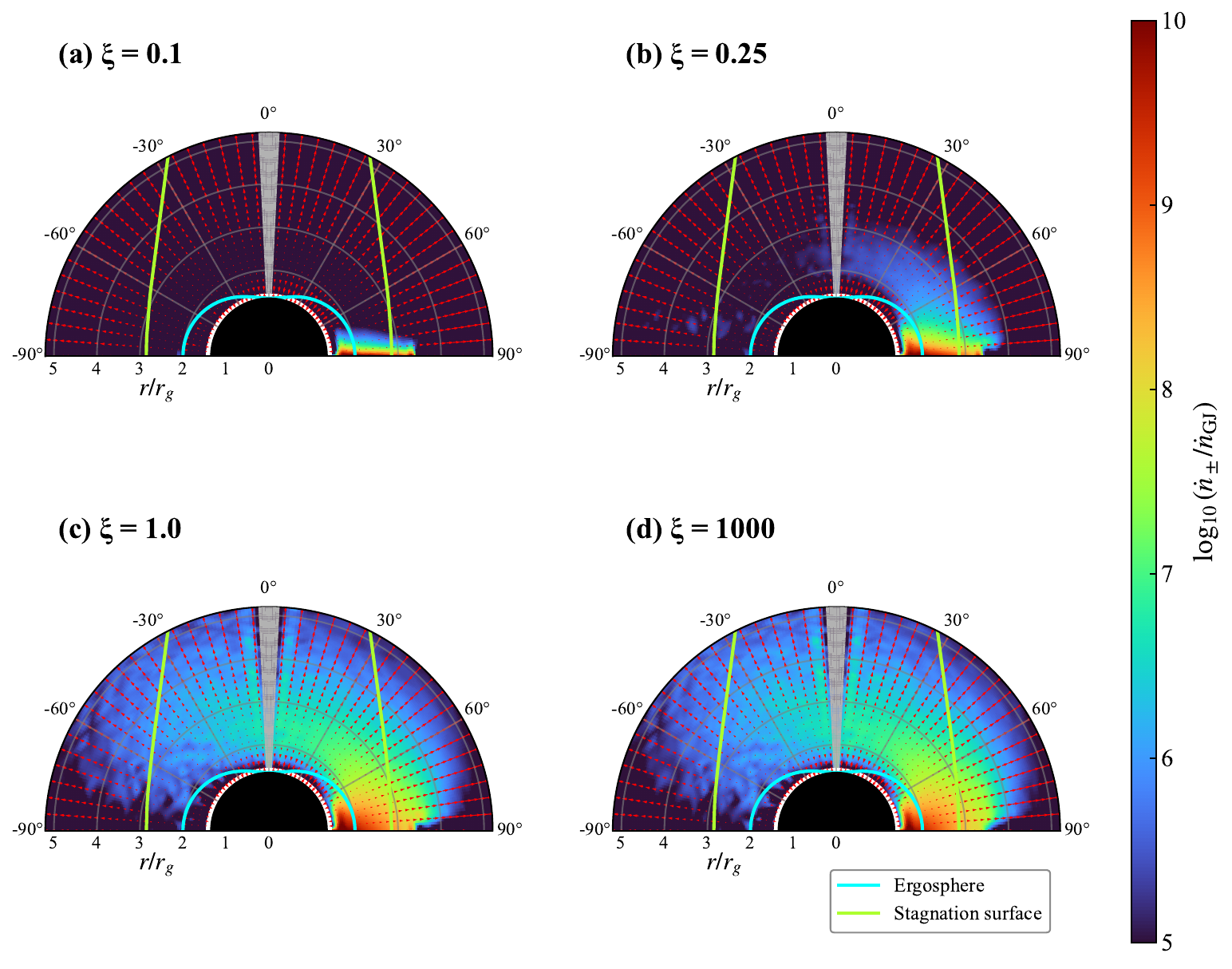}
\caption{Comparison of the spatial distribution of injected pairs on the meridional plane at $\varphi = 0$ for different degrees of anisotropy, $\xi = 0.1, 0.25, 1.0,$ and $1000$ (see Fig.~\ref{anisotropic function graph}), adopting a spin parameter $a = 0.9375$. The corresponding equatorial-plane distributions are shown in Section~\ref{Appendix:Spartial distribution in equator} (Fig.~\ref{Anisotropy2}).}
\label{Anisotropy1}
\end{figure*}

 In the Minkowski metric, where spacetime is flat, photon trajectories are not bent and do not wrap around the central object; as a result, pair production occurs only in the vicinity of the reconnection region. In contrast, in the Schwarzschild case, photon trajectories can bend around the black hole, allowing pairs to be produced on the opposite side of the reconnection region (see also the equatorial-plane distribution in Fig.~\ref{GR2}).
A particularly interesting result is obtained for spinning black holes: On the meridional plane, pairs are also injected near the rotation axis, i.e., the spine region. This feature is clearly seen in the polar-angle distribution of injected pairs at $r = 3\,r_g$ (see the middle panal of Fig.~\ref{GR_theta_distribution}), which shows that the injected pairs into the spine region in the Kerr black hole is enhanced by $\sim 10^3$--$10^5$ compared to the non-spinning black hole. This behavior originates from the fact that photon orbit in the Kerr metric is not confined to a fixed orbital plane, as the fundamental frequencies in the radial, polar ($\theta$), and azimuthal ($\varphi$) directions are generally incommensurate \citep{Fujita+09,van}. As a result, the fraction of head-on collisions near the rotation axis increases compared to the Schwarzschild case, enhancing pair injection into the spine region.
The same effect leads to a broader spatial distribution of pairs in the jet region as the spin parameter increases, and the pair production rate correspondingly becomes higher for larger spins (Table~\ref{Comparison with each spacetime}).

{The bottom panel of Fig.~\ref{GR_theta_distribution} shows the radial velocity of the injected pairs at $\theta=\pi/2$ in the drift frame. In the Kerr spacetime, the radial velocity is comparable to or exceeds the reconnection inflow velocity ($\tilde{\beta}_{\rm rec}\approx0.1$). For $a=0.9375$, this indicates that pairs injected near the stagnation surface possess sufficient outward momentum to escape and contribute to jet mass loading.}

{On the other hand, as shown in Fig.~\ref{GR1}, the pair injection region lies inside the stagnation surface for all cases except $a=0.9375$. Nevertheless, a amount of pairs sufficient to account for the radio emission are still injected at $r>r_h$ (see $\dot{N}_{r>r_h}$ in Table~\ref{Comparison with each spacetime}). These pairs may be further accelerated by their initial outward momentum and thermal pressure gradients and contribute to jet mass loading (see Section~\ref{Discussion:SSA}).}

\subsection{Influence of photon anisotropy}\label{Result:Anisotropic influence}

The results presented so far have primarily assumed isotropic photons ($\xi = 1000$); however, radiation from relativistic electrons accelerated by magnetic reconnection is intrinsically anisotropic due to the electromagnetic field structure and relativistic beaming effect \citep{Cerutti+14,Chernoglazov+23}. In this section, we examine the spatial distribution of injected pairs for four representative cases, $\xi = 0.1, 0.25, 1.0,$ and $1000$ (see Section~\ref{Model:photon field}), adopting a spin parameter $a = 0.9375$, in order to assess the impact of photon anisotropy. 

\begin{table}[t]
    \centering
    \vspace{20pt}
    \begin{tabular}{lccc}
        \hline\hline
        $\xi$
        & ${\sigma}$
        & $\dot{N}_{r>r_{\rm stag}}\;\left(\; \rm s^{-1}\right)$
        & $L_{\rm43GHz}\;\left(\rm erg\;s^{-1}\right)$
        \\
        \hline
        $0.1$ & $1.3\times10^3$ & $2.1\times10^{42}$ & $1.3\times10^{37}$ \\
        $0.25$ & $1.1\times10^4$ & $1.4\times10^{43}$ & $8.3\times10^{37}$ \\
        $1.0$ & $3.5\times10^4$ & $6.8\times10^{43}$ & $4.1\times10^{38}$ \\
        $1000$ & $5.6\times10^4$ & $6.0\times10^{43}$ & $3.5\times10^{38}$ \\
        \hline\hline
    \end{tabular}
    \caption{Comparison of the self-regulated upstream magnetization parameter of the reconnection region for $f_{\rm adv}=1$, the pair production rate, and the radio synchrotron emission at $r = 1000\,r_g$ at 43 GHz for different degrees of anisotropy. Within the parameter range considered here, the self-regulated magnetization satisfies $\sigma<\gamma_{\rm syn}$ in all the cases. In addition, the radio luminosity reaches $L_{43\,\mathrm{GHz}} \gtrsim 10^{37}\,\mathrm{erg\ s^{-1}}$, indicating that a sufficient amount of plasma is supplied to account for the observed radio emission from the M87* jet.}
    \label{Comparison with each anisotropy}
\end{table}

\begin{figure}[t]
\raggedright
\vspace{10pt}
\includegraphics[width=\columnwidth]{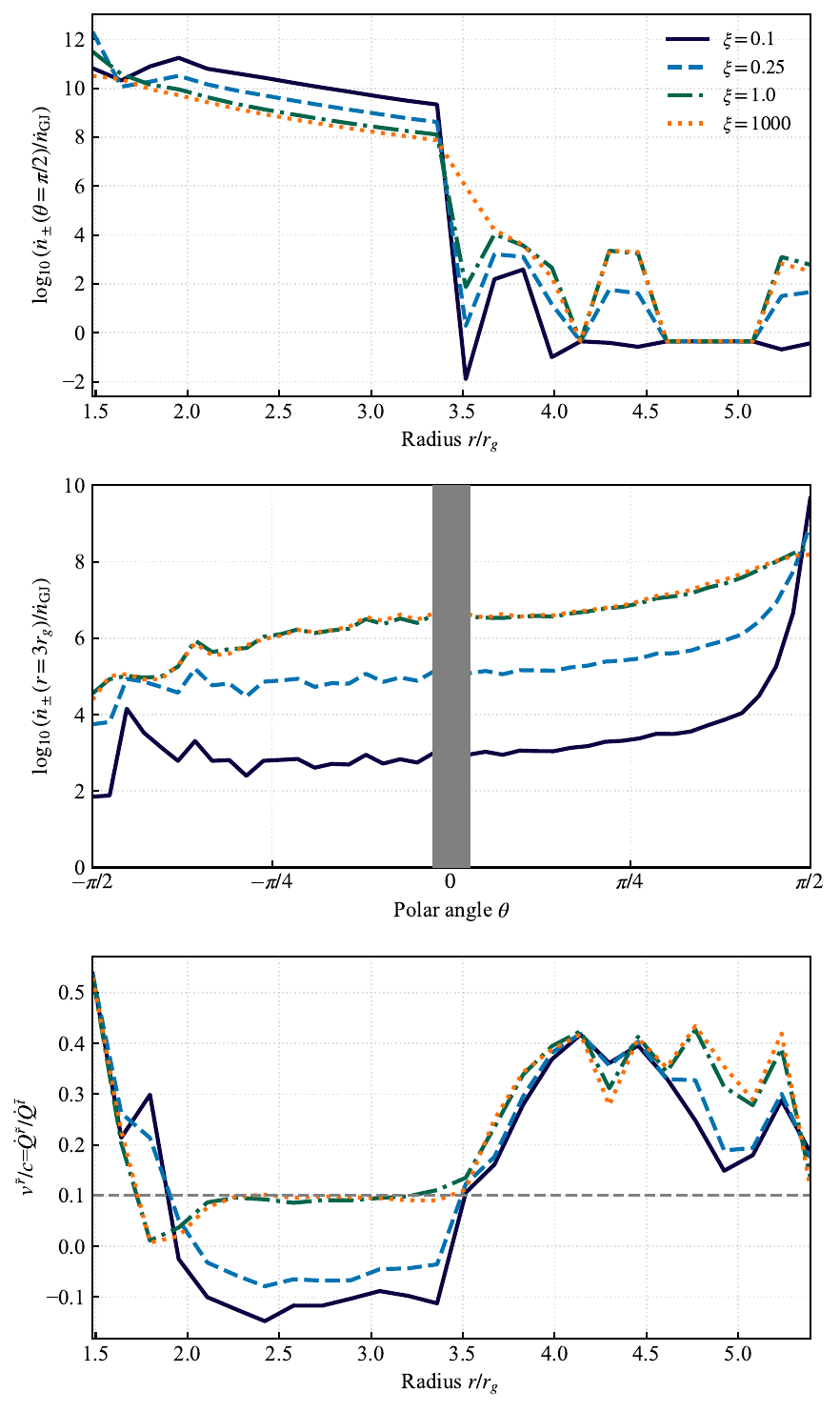}
\caption{{Pair production rate and velocity distributions obtained from Fig.~\ref{Anisotropy1}. The top, middle, and bottom panels show the same quantities as in Fig.~\ref{GR_theta_distribution}.}
}
\label{Anisotropy_theta_distribution}
\end{figure}

Within the parameter range adopted in this work, all cases fall within the weak cooling regime (see Table~\ref{Comparison with each anisotropy}), in which the photon distribution is preferentially concentrated along the direction of the reconnection electric field  (see Section~\ref{Model:photon field}). Fig.~\ref{Anisotropy1} shows the spatial distribution of injected pairs on the meridional plane at $\varphi = 0$ for different degrees of photon anisotropy. We find that the injected pairs are similarly concentrated near the reconnection region (see the top panel of Fig.~\ref{Anisotropy_theta_distribution}), while the anisotropy suppresses pair injection into the vicinity of the rotation axis even in the Kerr metric (see also the middle panel of Fig.~\ref{Anisotropy_theta_distribution}). This behavior arises because the reconnection electric field is oriented predominantly within the equatorial plane; consequently, increasing anisotropy leads to a progressively stronger concentration of photons along directions parallel to the equatorial plane. For such a photon distribution, a large fraction of photons satisfy the conserved quantity of geodesic motion, the Carter constant $Q \simeq 0$, which suppresses oscillatory motion in the $\theta$ direction. Consequently, photon trajectories are confined near the equatorial plane, leading to a suppression of pair production in the vicinity of the black hole rotation axis. Therefore, in contrast to the isotropic case, stronger anisotropy confines pair production to a limited region just above the magnetic reconnection site, rather than producing a spatial distribution that extends throughout the entire jet region. 

We note that the pair multiplicity remains larger than unity in the jet region, i.e., $\dot{n}_{\pm}/\dot{n}_{\rm GJ} > 1$, as shown in Fig.~\ref{Anisotropy_theta_distribution},  although the plasma density near the rotation axis is reduced by up to $\sim 3$ orders of magnitude compared to the isotropic case. (See also Appendix~\ref{Appendix:spartial distribution} and Fig.~\ref{M87_kappa_appendix} for the full dynamic range of the injected pair number density.) 
In addition, the anisotropy reduces the total pair production rate within the jet (see Table~\ref{Comparison with each anisotropy}) because the injected pairs no longer populate the entire jet region. Nevertheless, our scenario still supplies a sufficient amount of plasma to account for the observed radio emission from the jet, even in strongly anisotropic cases. {However, the bottom panel of Fig.~\ref{Anisotropy_theta_distribution} shows that the radial velocity of the injected pairs becomes inward in highly anisotropic cases when measured in the drift frame, implying that their contribution to jet mass loading may be reduced compared to the isotropic case. To investigate how much of the injected pairs in highly anisotropic cases contribute to jet mass loading requires self-consistent GRMHD simulations that incorporate pair injection physics.}

\section{Discussion} \label{sec:Discussion}

\subsection{Expected Degree of Photon Anisotropy in M87*}


{We adopt the energy-independent anisotropy model given by Eq.~\eqref{anisotropic function} for simplicity. This treatment seems inconsistent with recent PIC simulations that suggest  energy-dependent photon anisotropy \citep{Cerutti+14,Chernoglazov+23}. Nevertheless, we consider that our approximation is reasonable as described below. Fig.~\ref{injection spectral} shows that photons with energies well below the burn-off limit contribute comparably to the pair injection rate. According to Fig.~14 of \citet{Chernoglazov+23}, these lower-energy photons become nearly isotropic in the regime relevant to our model, $\sigma/\gamma_{\rm syn}\sim0.01$. Therefore, although some degree of anisotropy is expected to remain, its overall impact is likely moderate, and we expect the results obtained with $\xi=1.0$ to be representative of M87*. In addition, preliminary results from our ongoing work indicate that the degrees of anisotropy are moderate also for most of other jet-producing AGNs.}

{As discussed in Section~\ref{Result:Anisotropic influence}, when the photon distribution is nearly isotropic, pair injection occurs over a much broader region, supplying a substantial number of pairs throughout the jet. Such a spatial distribution has important implications for jet acceleration (Section~\ref{Discussion:SSA}) and the formation of magnetospheric spark gaps (Section~\ref{Discussion:VHE}).}

\subsection{Jet Acceleration by pressure gradient} \label{Discussion:SSA}
In Section~\ref{Resuly:GR influence}, we demonstrated that a spinning black hole plays a key role in facilitating pair injection across the entire jet region, as seen in the spatial distribution of Fig.~\ref{GR1}. The injected plasma emits synchrotron radiation in the magnetosphere; however, the plasma may not cool efficiently and could instead become thermalized due to the high synchrotron self-absorption (SSA) opacity \citep{Kimura&Toma22}. {This thermalization is primarily driven by the absorption of low-energy synchrotron photons \footnote{For the parameters of M87*, the synchrotron photons emitted by the injected pairs are primarily in the infrared band \citep{Kimura&Toma22,Hokobyan+23}.}.} Injected pairs inside the stagnation surface—defined as the surface where the Lorentz force balances gravity—can be accelerated via adiabatic expansion by the thermal pressure gradient.
Here, based on the results of our numerical calculations, we examine the thermalization of the injected pairs and its implication on the bulk acceleration.
\begin{figure}[t] 
\includegraphics[width=\columnwidth]{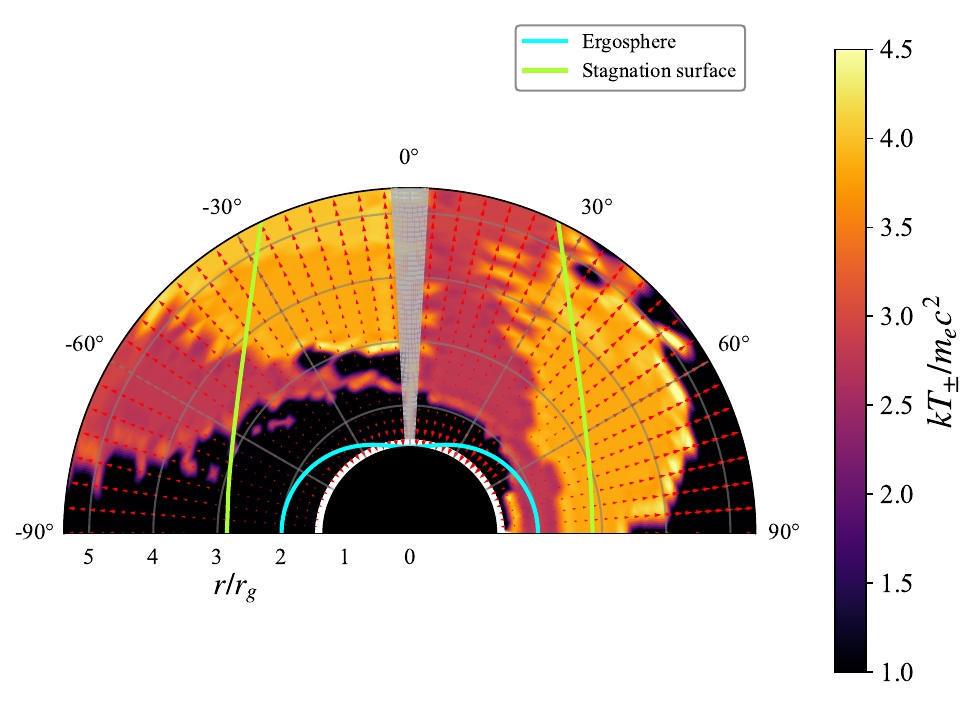} \caption{Temperature distribution of SSA-thermalized injected plasma in M87* ($a = 0.9375$). The distribution is shown only in regions satisfying $t_{\rm SSA} < t_{\rm \pm,\, syn}<t_{\rm dyn}$. This result suggests that even the plasma inside the stagnation surface may be accelerated to relativistic speeds by thermal pressure gradients.} 
\label{Discussion:temparature_distribution} 
\end{figure} 

\begin{figure*}[t] 
\includegraphics[width=\textwidth]{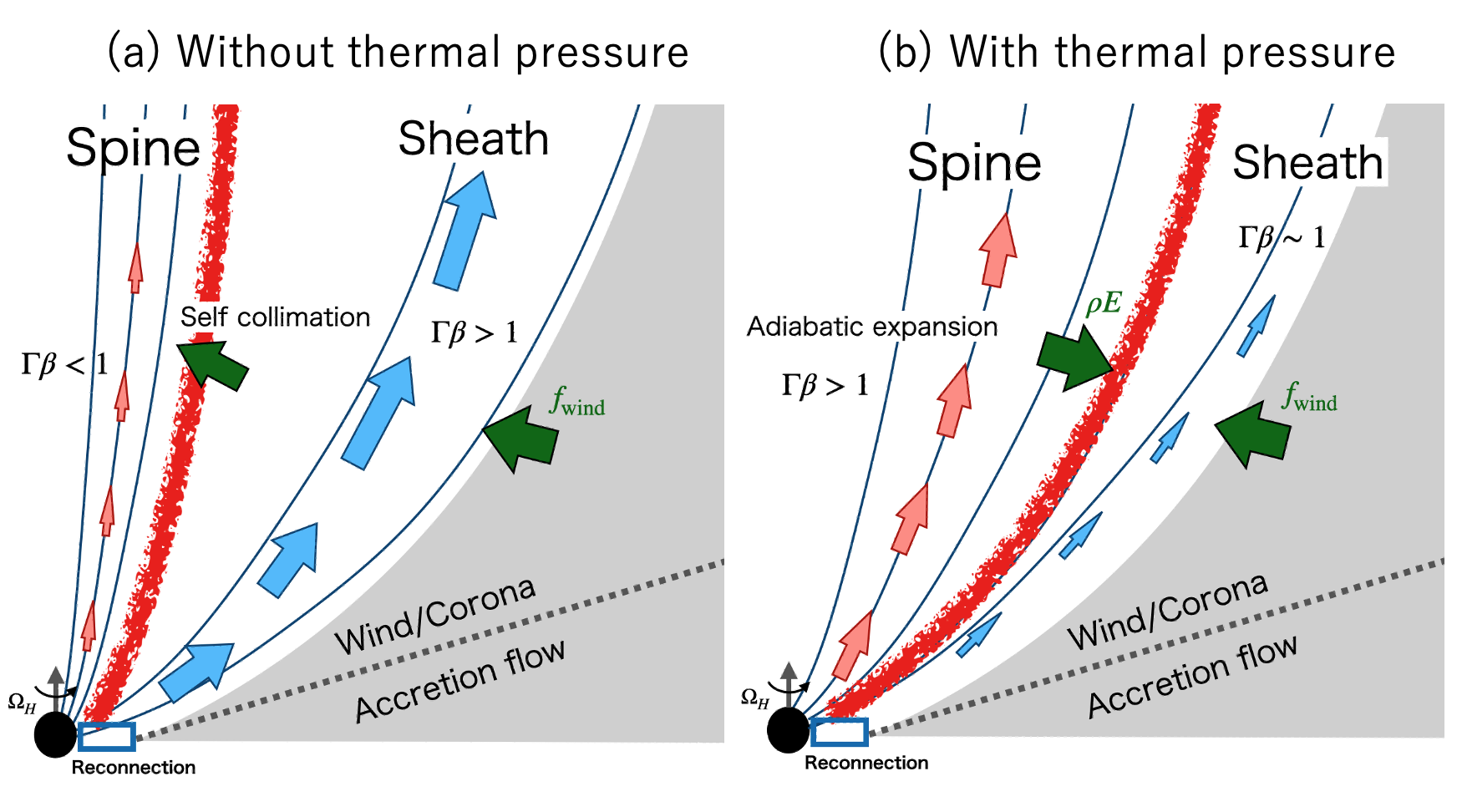} \caption{Schematic picture of the jet acceleration in the cold MHD (left panel; (a)) and that suggested by the spatial distribution of injected pairs in this work (right panel; (b)). In the cold MHD scenario, there is little current crossing magnetic field lines near the jet spine, making Lorentz-force acceleration inefficient and resulting in a non-relativistic flow and magnetic self-collimation near the jet spine. Since the jet is collimated by external pressure and the spine region is further compressed, the magnetic field configuration in the jet sheath expands more widely than in the spherical case, allowing the sheath to be efficiently accelerated.
In contrast, our matter-loading scenario supplies plasma near the jet spine in the Kerr metric, accelerating it to relativistic speeds via thermal pressure gradients. {As a result, the ideal-MHD electric field (${\bf E} = -{\bf v} \times {\bf B}/c$, where ${\bf v}$ is the jet velocity in the spine region) becomes dynamically important, particularly in the jet-spine region.} This effect leads to a more narrowly collimated jet sheath. Consequently, the efficiency of magnetic-to-kinetic energy conversion in the jet sheath is reduced, suggesting that its terminal velocity may be lower than that predicted by the cold-MHD picture.}
\label{Discussion:SSA fireball picture} 
\end{figure*}

The SSA heating timescale is given by \citep{Asano2001, Kimura&Toma22}
\begin{equation}
    t_{\rm SSA} = \frac{\gamma_{\pm} m_e c^2}{\int dE_{\gamma}\, E_{\gamma}\, n_{E_{\gamma}}\, \sigma_{\rm SSA}\, c},
\end{equation}
where $\sigma_{\rm SSA}$ is the SSA cross section \citep{Ghisellini1991}, 
$\gamma_{\pm} = E_{\pm}/(m_e c^2)$ is the Lorentz factor of the injected pairs, 
and $E_{\gamma} n_{E_{\gamma}}$ represents the photon energy density emitted by the injected plasma. 
Following the calculation in Appendix E of \citet{Kimura&Toma22}, we obtain
\begin{equation}
    t_{\rm SSA} \approx \frac{12\pi}{n_{\pm} c\, \sigma_{\rm SSA,0}}
    \max\left(1, \frac{t_{\rm \pm,\,syn}}{t_{\rm dyn}}\right)\left(\frac{\nu_{\rm min}}{\nu_{\rm max}}\right)^{-4/3},
\end{equation}
where
\begin{align}
t_{\rm dyn} &= \frac{l_{\rm rec}}{c}, 
\quad 
t_{\rm \pm,\,syn} = \frac{6\pi m_e c}{\sigma_T B^2 \gamma_{\pm}}, 
\end{align}
$\sigma_{\rm SSA,0} = 2^{11/3}3^{7/6}\pi^2 \Gamma^2\!\left(\tfrac{4}{3}\right)e/(5B)$, $\nu_{\rm min} = {eB}/{(2\pi m_e c)}$ and $\nu_{\rm max} = {eB\gamma_{\pm}^2}/{(2\pi m_e c)}$. Here $\Gamma(x)$ denotes the Gamma function. The term of $\max\left(1,{t_{\rm \pm,\,syn}}/{t_{\rm dyn}}\right)$ represents the effect of synchrotron cooling of the injected pairs. For the case where $t_{\rm SSA} < t_{\pm,\,\rm syn} < t_{\rm dyn}$, the injected pairs are efficiently thermalized without significant energy loss. 
Fig.~\ref{Discussion:temparature_distribution} shows the resulting temperature distribution of the injected plasma that is efficiently thermalized in M87*. The temperature is evaluated by first computing the average energy of electron–positron pairs ${\langle \hat{E}_{\pm}\rangle}$ at the local pair-production site, and then imposing particle-number and energy conservation during the thermalization process to obtain the post-thermalization temperature ($kT_{e}=\langle \hat{E}_{\pm}\rangle/3$). We find that the electron–positron pairs located inside the stagnation surface are thermalized via the SSA process and the temperature satisfies $kT_{e} > m_e c^2$, indicating that the mean lorentz factor of the thermalized pairs is $\langle{\hat{\gamma}}_{\pm}\rangle\approx3kT_e/(m_ec^2)\sim10$. This value is larger than 
the estimate under the assumption of head-on collisions of photons in \citet{Kimura&Toma22} by a factor of a few. This difference arises because the collisions of higher energy photons with smaller angles contribute to the pair creation.
{These results suggest that the reconnection region may be dominated by hot plasma, which could potentially affect the reconnection dynamics. 
However, the detailed influence of hot pairs on the overall reconnection dynamics remains poorly understood and warrants further investigation.}

Such high-temperature plasma can be accelerated to relativistic speeds by the thermal pressure gradient even when located inside the stagnation surface. This result is also different from the conventional picture of MHD jet acceleration. MHD simulation suggests that there is little current crossing magnetic field lines near the jet spine, making Lorentz-force acceleration inefficient and leading to a non-relativistic flow and magnetic self-collimation near the axis. The whole jet is collimated by the external pressure from the disk wind, leading to a magnetic field configuration in the sheath that expands more widely than the spherical profile toward larger radii, with efficient acceleration occurring (see panel~(a) of Fig.~\ref{Discussion:SSA fireball picture}) \citep{komissarov2009,Lyubarsky+09,Nakamura2018}.
In contrast, our results show that $e^{\pm}$ pairs can be injected into the vicinity of the black hole rotation axis in Kerr spacetime, then, these pairs are efficiently thermalized via SSA and can be accelerated to relativistic speeds by the thermal pressure gradient. If relativistic plasma flows are formed even near the jet spine, the contribution of the outward electric field stress can no longer be neglected. As a result, magnetic self-collimation becomes less efficient, and it becomes difficult to form a magnetic field configuration that expands significantly beyond the spherical profile at the sheath
(see Fig.~\ref{Discussion:SSA fireball picture}, panel~(b)). This implies that the efficiency of converting electromagnetic energy (or thermal energy) into plasma kinetic energy via the magnetic pressure gradient of the toroidal field (or thermal gradiant pressure) may be ineffective in the jet sheath. This is likely consistent with observations of active galactic nucleus jets, which suggest that jet acceleration is slower than that predicted by ideal MHD \citep{Nakamura2018,Park+2019a}.

Our results indicate that the discrepancy between theoretical models of jet acceleration and observations could be explained by the thermal pressure gradient of plasma near the spine, which has not been fully considered in previous studies.
The validity of this picture should be confirmed by numerical simulations of acceleration of injected matter.

\subsection{Black hole magnetospheric dynamics}\label{Discussion:VHE}

{Our matter-loading scenario has direct implications for the dynamics of the black-hole magnetosphere. As shown in Section~\ref{sec:result}, the pair density produced by magnetic reconnection exceeds the Goldreich--Julian density throughout the jet (see Fig.~\ref{M87_GR_appendix}), even when anisotropic photon distributions are considered (see Fig.~\ref{M87_kappa_appendix}). The injected pairs are therefore expected to supply the electric current required to sustain the BZ process and efficiently screen the electric field component parallel to the magnetic field. This suggests that the magnetosphere is unlikely to become sufficiently charge-starved for a spark gap to form.}

{Nevertheless, the results of \citet{Enzo2026} suggest a possible coexistence of our matter-loading scenario and intermittent spark-gap formation. Although their study is based on a 2D global GRPIC simulation of Bondi accretion onto a Schwarzschild black hole, they identify three stages that repeat cyclically in the MAD state: (1) an ideal advection phase, (2) a reconnection-regulated phase, and (3) a large-scale reconnection phase similar to that reported by \citet{Ripperda22}. Our matter-loading mechanism primarily operates during phases (2) and (3), supplying plasma to the BZ jet. In these phases, the electric field is expected to be efficiently screened throughout the jet region, as discussed above.}

{In contrast, during (1) the advection phase, magnetic reconnection is not efficiently driven and our matter-loading mechanism becomes ineffective. As a result, the magnetosphere near the black hole may temporarily transition to a low-density state owing to plasma outflow into the jet and inflow into the black hole. In this regime, the plasma density can fall below the Goldreich--Julian density, allowing spark gaps to form. Such intermittent gap formation may be relevant to the origin of very-high-energy ($300\,{\rm GeV}$--$10\,{\rm TeV}$) flares observed in systems such as M87* \citep{EHT2024}. These flares exhibit rapid variability on timescales of days, suggesting a compact emission region located near the black hole, i.e., at the jet base \citep{Hada2024}. Particle acceleration in spark gaps within BZ jets has long been proposed as a possible mechanism for producing TeV emission \footnote{Another scenario for variable TeV emission is inverse Compton emission associated with large-scale magnetic reconnection in the MAD state \citep{Ripperda22,Hokobyan+23,Solanki+25}. However, our results suggest a self-regulated magnetization parameter of $\sigma \sim 10^5$, which may make the production of TeV photons through inverse Compton scattering challenging \citep{Chne+23}. As discussed by \citet{Solanki+25}, anisotropic inverse Compton emission and current sheets located away from the equatorial plane may play an important role in overcoming this difficulty.}\citep{Levinson2011L,Hirotani+16,Crinquand+20,Kisaka+22}. The cyclic transitions reported by \citet{Enzo2026} may therefore provide a natural framework for intermittently activating such gaps, potentially giving rise to the rapid variability of TeV flares.}

{It should be noted, however, that the results of \citet{Enzo2026} are based on 2D simulations of a non-rotating black hole, and it remains unclear whether these three phases persist in fully 3D systems of a rotating black hole. Future global 3D GRPIC simulations will be necessary to verify the existence of these phases and to assess their role in magnetospheric plasma supply, spark-gap formation, and high-energy emission.}

\section{Summary}
In this work, we have derived the spatial distribution of injected plasma by calculating the $e^{\pm}$ pair production from high-energy photons originating from non-axisymmetric magnetic reconnection, as suggested by recent 3D GRMHD simulations. Our calculation employs general relativistic ray tracing, taking into account curved spacetime, photon-field anisotropy, and collision angles. 

We find that pair injection driven by non-axisymmetric magnetic reconnection can supply a sufficient number of particles to account for the observed jet emission as claimed by \citet{Kimura&Toma22}. The self-regulated magnetization achieved through pair creation is consistent with the values reported in \citet{Chne+23,Kuze+24}. {We show that pair production is dominated by synchrotron photons emitted by electrons in the high-energy power-law tail ($\sigma<\tilde{\gamma}<\gamma_{\rm syn}$).} A key new result of this study is that this conclusion remains valid even when the anisotropy of the photon field is taken into account, thereby establishing a viable plasma source for the jet. We also demonstrate a new finding of this study: {The black hole spacetime has a strong impact on the spatial distribution of injected pairs.} Owing to frame-dragging effects, photon trajectories are no longer confined to a single plane, which enhances head-on collisions in the spine region and leads to efficient pair injection. As a result, relativistic outflows can be formed in the spine via the thermalization by SSA and the adiabatic expansion while energy conversion in the sheath will become less efficient. This may lead to slower sheath flows than those typically found in MHD simulations, potentially bringing theoretical predictions into better agreement with observations. In addition, we find that the spatial distribution of the injected plasma can significantly influence black hole magnetospheric dynamics.

In this study, we have evaluated the pair production rate assuming the mass and accretion rate of M87*. It is important to investigate whether this matter-loading scenario can be applied to other black hole jet systems. A systematic parameter survey over black hole mass and accretion rate will therefore be essential in separate work. Furthermore, during thermal acceleration in the spine, inverse Compton cooling and adiabatic cooling may become important, and it remains unclear to what extent the spine–sheath structure can be maintained within the jet. Since the formation of a spine–sheath structure is crucial for understanding relativistic outflows in blazars, this issue warrants further investigation. Incorporating thermal pressure acceleration due to the SSA process into MHD simulations will be an important direction for future work.

\begin{acknowledgments}
We would like to thank Tomohisa Kawashima, Koki Kin, Shota Kisaka, Riku Kuze, and Kengo Tomida for their useful comments. We also appreciate useful comments of anonymous referee. The numerical computations in this study are performed on the XD2000 at the National Astronomical Observatory of Japan, and Genkai at Kyushu University. This work was supported by Graduate Program on Physics for the Universe (GP-PU).
\end{acknowledgments}



\appendix

\section{The Kerr spacetime and the 3+1 electrodynamics}\label{Appendix:Kerr_spacetime_3+1electrodynamics}
We use Boyer–Lindquist coordinates $(t,\varphi,r,\theta)$ and adopt units in which $GM=c=1$ (so that $r_g = GM/c^2 = 1$). The Kerr spacetime \citep{Kerr63} can be written as
\begin{equation}
ds^2 = g_{\mu\nu}dx^{\mu}dx^{\nu} =- \alpha^2 dt^2 + \gamma_{\varphi\varphi}(d\varphi-\Omega dt)^2+\gamma_{rr}dr^2 + \gamma_{\theta\theta}d\theta^2,
\label{metric}
\end{equation}
where 
\begin{equation}
\alpha = \sqrt{\frac{\rho^2\Delta}{\Sigma}},\quad
\Omega = \frac{2ar}{\Sigma}, \quad
\gamma_{\varphi\varphi} = \frac{\Sigma \sin^2\theta}{\rho^2}, \quad
\gamma_{rr} = \frac{\rho^2}{\Delta}, \quad
\gamma_{\theta\theta} = \rho^2.
\end{equation}
Here, we have defined $\Sigma=(r^2+a^2)^2-a^2\Delta\sin^{2}{\theta}$, $\Delta=r^2+a^2-2r$, $\rho^2=r^2+a^2\cos^2{\theta}$, where $a$ denotes the dimensionless spin parameter. The Kerr spacetime admits two Killing vectors associated with stationarity and axisymmetry, 
\begin{equation}
    \xi^{\mu}_{(t)}=(1,0,0,0), \quad \xi^{\mu}_{(\varphi)}=(0,1,0,0).
\end{equation}

In Boyer--Lindquist coordinates, the four-velocity of an observer whose worldline is normal to hypersurfaces of constant $t$ is given by
\begin{equation}
    n^{\mu}=\left(\frac{1}{\alpha},\frac{\Omega}{\alpha},0,0\right), 
    \qquad 
    n_{\mu}=(-\alpha,0,0,0).
\end{equation}
This observer is referred to as a zero-angular-momentum observer (ZAMO), since $n_{\mu}\xi^{\mu}_{(\varphi)}=0$. The ZAMO has a local orthonoramal basis \citep{Throne1986,Komissarov04,Toma2014},
\begin{equation}
    e_{(\hat{t})}=\frac{\partial_t+\Omega\partial_{\varphi}}{\alpha},\quad e_{(\hat{\varphi})}=\frac{\partial_{\varphi}}{\sqrt{\gamma_{\varphi\varphi}}},\quad 
    e_{(\hat{r})}=\frac{\partial_{r}}{\sqrt{\gamma_{rr}}}, \quad
    e_{(\hat{\theta})}=\frac{\partial_{\theta}}{\sqrt{\gamma_{\theta\theta}}},
\end{equation}
\begin{equation}
    w^{(\hat{t})}=\alpha dt,\quad w^{(\hat{\varphi})}=\sqrt{\gamma_{\varphi\varphi}}(d\varphi-\Omega dt),\quad 
    w^{(\hat{r})}=\sqrt{\gamma_{rr}}dr, \quad
    w^{(\hat{\theta})}={\sqrt{\gamma_{\theta\theta}}}d\theta.
\end{equation}
Physical quantities expressed in the ZAMO orthonormal basis are denoted by a hat, e.g., $A^{\hat{\mu}}$ or $\hat{A}$.

The covariant Maxwell equations
$\nabla_{\nu}F^{\mu\nu}=4\pi I^{\mu}$, 
$\nabla_{\nu}{}^{*}F^{\mu\nu}=0$ 
can be reduced by introducing the four vectors
$D^{\mu}=n_{\nu}F^{\mu\nu}$,
$B^{\mu}=-n_{\nu}{}^{*}F^{\mu\nu}$,
$E_{\mu}=\xi_{(t)}^{\nu}F_{\mu\nu}$, and
$H_{\mu}=-\xi_{(t)}^{\nu}{}^{*}F_{\mu\nu}$ to \cite{Laudau1975,Komissarov04} 
\begin{equation}
    \nabla \cdot \bm{B}=0,\quad\frac{\partial \bm{B}}{\partial t}+\nabla \times \bm{E}=0, \label{maxwell eq1}
\end{equation}
\begin{equation}
    \nabla \cdot \bm{D}=4\pi \rho_e,\quad -\frac{\partial \bm{D}}{\partial t}+\nabla \times \bm{H}=4\pi \bm{J}, \label{maxwell eq2}
\end{equation}
where, $\rho_e=-I^{\mu}n_{\mu}$ and $J^{\mu}=(\xi_{(t)}^{\mu}I^{\nu}-\xi_{(t)}^{\nu}I^{\mu})n_{\nu}$. $\nabla \cdot \bm{A}$ and $\nabla \times \bm{A}$ denote $(1/\sqrt{\gamma})\partial_{i}(\sqrt{\gamma}A^{i})$ and $(1/\sqrt{\gamma})\varepsilon^{ijk}\partial_{j}A_k$, and $\varepsilon^{ijk}$ is the Levi-Civita pseudo-tensor. $\bm D$ and $\bm B$ are the electric and magnetic fields measured by ZAMO, while $\bm E$ and $\bm H$ are those in the coordinate basis, and they have constitutive relations,
\begin{equation}
    \bm{E}=\alpha\bm{D}-\Omega\bm{m}\times \bm{B},\quad\bm{H}=\alpha \bm{B}+\Omega \bm{m}\times \bm{D}, \label{dual relation}
\end{equation}
where $\bm{m}=\partial_{\varphi}$.

\section{Electromagnetic field model around black hole}\label{Appendix:Electromagnetic_field_model}
 High resolution GRMHD simulation by \cite{Ripperda22} suggests that the accretion disk in the MAD becomes geometrically thin and the magnetic field configuration approaches a split-monopole geometry within $r \lesssim 5r_g$. The force-free condition provides a good approximation between the inner and light surfaces. In this paper, we take the assumptions of stationarity, axisymmetry, a force-free condition ($F_{\mu\nu} I^{\nu}=0$), and a split-monopole magnetic field configuration ($B^{\theta}=0$), and we model electomagnetic field around the black hole in the same manner as \cite{Kimura&Toma22}.

Under the assumptions of force-free conditions, stationarity, and axisymmetry, the electric fields $\bm{E}$ and $\bm{D}$ can be expressed, using Equations~(\ref{maxwell eq1}) and (\ref{dual relation}), as
\begin{equation}
\bm{E} = -\Omega_{F}\bm{m} \times \bm{B},
\qquad
\bm{D} = \frac{1}{\alpha} (\Omega - \Omega_F)\bm{m} \times \bm{B}, \label{electric field}
\end{equation}
where $\Omega_F$ denotes the angular velocity of magnetic field lines induced by the spinning black hole ($a \neq 0$); notably, $\Omega_F$ is conserved along each field line. In this work, we adopt $\Omega_F = a / 4 r_{h}$ \citep{Blandford&Znajek77}, where $r_h = r_g \left( 1 + \sqrt{1 - a^2}\right)$ is the horizon radius. 

For the non-zero components of the magnetic field, assuming a split-monopole configuration as well as stationarity and axisymmetry, Eqs.~\eqref{maxwell eq1} and \eqref{maxwell eq2} lead to the existence of conserved quantities along magnetic field lines:
\begin{equation}
\sqrt{\gamma} B^r = \mathrm{const},
\qquad
H_{\varphi} = \alpha B_{\varphi} = \mathrm{const}.
\end{equation}
Using the ZAMO orthonormal basis
($B_{\hat{r}}=\sqrt{\gamma_{rr}}B^r,\; B_{\hat{\varphi}}=B_{\varphi}/\sqrt{\gamma_{\varphi\varphi}}$), we obtain
\begin{equation}
\frac{B_{\hat{\varphi}}}{B_{\hat{r}}} \propto \frac{\sqrt{\gamma_{\varphi\varphi}}}{\alpha}
\end{equation}
for a fixed $\theta$.
Analytical studies of special relativistic MHD flows in magnetically dominated regions \citep{Beskin10,Toma&Takahara13} have shown that the magnetic field components satisfy
\begin{equation}
B_{\hat{\varphi}} \approx - B_{\hat{r}}
\end{equation}
at the outer light surface.
Under this condition we obtain the radial dependences of the field components as shown in Fig 5 of \citet{Kimura&Toma22}. It is roughly $B_{\hat{r}} \sim B_{\hat{\varphi}} \propto r^{-2}$.

When the accretion flow is in the MAD state, the magnetic field strength on the horizon in the ZAMO basis is approximately given by $\hat{B}_{\rm mad}\approx\sqrt{\dot{M}c\Phi_{\rm mad}^2/16\pi^2r_h^2}$ \citep{Yuan&Narayan14,Kimura&Toma22}, where $\Phi_{\rm mad}\approx 50$ is the saturated
magnetic flux \citep{Tchekhovskoy11,Narayan12,Mckinney12}, and $\dot{M}$ is a mass accretion rate. Therefore, by normalizing the field such that the radial magnetic field strength on the horizon becomes $\hat{B}_{\rm mad}$, we obtain
\begin{equation}
B_{\hat{r}}=\frac{2 r_h r_g}{\sqrt{\Sigma}} \hat{B}_{\rm mad}.
\end{equation}
We adopt this solution as the electromagnetic field configuration prior to the onset of magnetic reconnection.

\section{The orthonormal basis vectors of the drift frame}\label{Appendix:orthonormal_basis}

The four-velocity of the drift frame is given by
\begin{equation}
    u^{\mu}=u^{t}\left(1,\,
    \Omega-\frac{\alpha}{\sqrt{\gamma_{\varphi\varphi}}}
    \frac{D_{\hat{\theta}}B_{\hat{r}}}{\hat{B}^2},\,
    \frac{\alpha}{\sqrt{\gamma_{rr}}}
    \frac{D_{\hat{\theta}}B_{\hat{\varphi}}}{\hat{B}^2},\,
    0\right).
\end{equation}
The orthonormal basis vectors of the drift frame are constructed as 
\citep{Krolik05,Beckwith08,Shcherbakov11,Dextor16}
\begin{eqnarray}
     e_{(\tilde{t})}^{\mu} &=& u^\mu , \label{fluid t component} \\
     e_{(\tilde{\varphi})}^{\mu} &=& (u_{\varphi}, 0, 0, -u_t)/N_{\varphi}, \label{fluid phi component} \\
     e_{(\tilde{r})}^{\mu} &=& (u_r u^t, -(u_t u^t + u_{\varphi}u^{\varphi}), 0, u_r u^{\varphi})/N_r , \label{fluid r component} \\
     e_{(\tilde{\theta})}^{\mu} &=& (u_\theta u^t,  u_{\theta}u^r,1+u_{\theta}u^{\theta}, u_\theta u^{\varphi})/N_\theta ,
     \label{fluid theta component}
\end{eqnarray}
where
\begin{eqnarray}
     N_r ^2 &=& -\gamma_{rr}(u_t u^t + u_{\varphi}u^{\varphi})(1+u_{\theta}u^{\theta}), \\
     N_{\theta}^2 &=& \gamma_{\theta\theta}(1+u_{\theta}u^{\theta}), \\
     N_{\varphi}^2 &=& -(u_t u^{t} + u_{\varphi}u^{\varphi})\Delta \sin^2{\theta}.
\end{eqnarray}

Using Eqs.~\eqref{fluid t component}--\eqref{fluid theta component}, 
the electromagnetic fields in the drift frame basis are written as
\begin{eqnarray}
  B_{\tilde{\varphi}} &=& e_{(\tilde{r})}^{\alpha} e_{(\tilde{\theta})}^{{\beta}} F_{\alpha {\beta}} \
  = B_{\hat{\varphi}}
  - \hat{\gamma} v_d^{\hat{r}} D_{\hat{\theta}}
  + \frac{\hat{\gamma}^2}{\hat{\gamma} + 1} v_d^{\hat{r}}
  \left( v_d^{\hat{r}} B_{\hat{\varphi}} - v_d^{\hat{\varphi}} B_{\hat{r}} \right),
  \label{fluid phi B field} \\
  B_{\tilde{r}} &=& e_{(\tilde{\theta})}^{\alpha} e_{(\tilde{\varphi})}^{{\beta}} F_{\alpha {\beta}}
  = B_{\hat{r}}
  + \hat{\gamma} v_d^{\hat{\varphi}} D_{\hat{\theta}}
  + \frac{\hat{\gamma}^2}{\hat{\gamma} + 1} v_d^{\hat{\varphi}}
  \left( v_d^{\hat{\varphi}} B_{\hat{r}} - v_d^{\hat{r}} B_{\hat{\varphi}} \right),
  \label{fluid r B field} \\
  B_{\tilde{\theta}} &=& e_{(\tilde{\varphi})}^{\alpha} e_{(\tilde{r})}^{{\beta}} F_{\alpha{\beta}}
  = 0,
  \label{fluid theta B field} \\
  D_{\tilde{i}} &=& e_{(\tilde{t})}^{\alpha} e_{(\tilde{i})}^{{\beta}} F_{\alpha {\beta}}
  = 0,
  \label{fluid theta D field}
\end{eqnarray}
where $\hat{\gamma}=\left(1-\hat{\bm{v}_d}^2\right)^{-1/2}$ is the Lorentz factor of the drift frame measured by the ZAMO.
These expressions explicitly demonstrate that the electric field vanishes 
locally in the drift frame (Eq.~\eqref{fluid theta D field}). 
Moreover, Eqs.~\eqref{fluid phi B field} and \eqref{fluid r B field} show that 
the transformation between the ZAMO basis and the drift frame basis 
reduces exactly to the Lorentz transformation.

\section{Spatial Distribution of Pair-Injected Plasma in the Equatorial Plane}\label{Appendix:Spartial distribution in equator}
Here, we show the spatial distribution of pair-injected plasma on the equatorial plane corresponding to the results discussed in Sections~\ref{Resuly:GR influence} and \ref{Result:Anisotropic influence}.

\begin{figure*}[t]
\centering
\includegraphics[width=\textwidth]{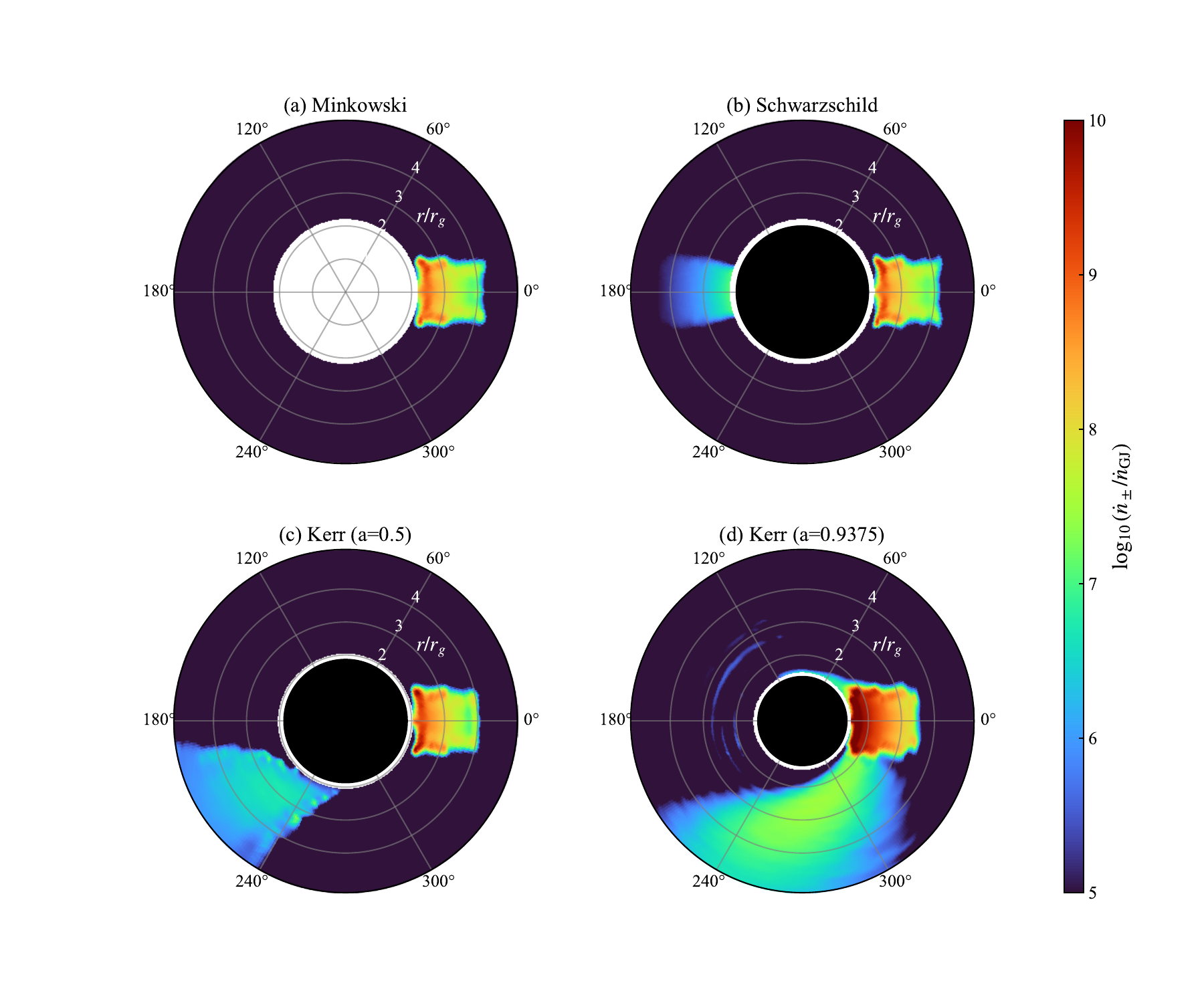}
\caption{Comparison of the spatial distribution of pair injection into the magnetosphere on the equatorial plane for different spacetime geometries, corresponding to the equatorial-plane of Fig.~\ref{GR1}.}
\label{GR2}
\end{figure*}

\begin{figure*}[t]
\centering
\includegraphics[width=0.82\textwidth]{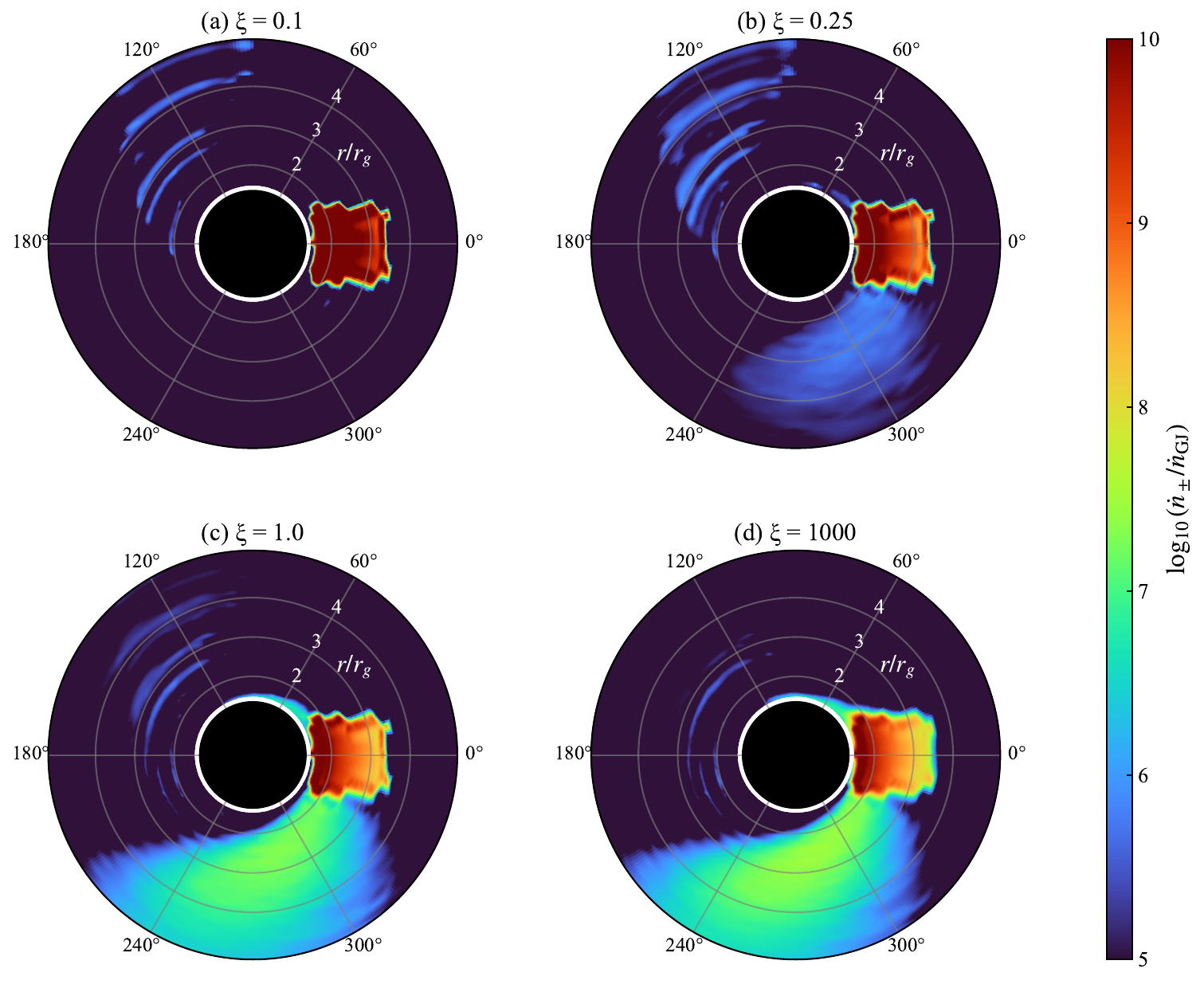}
\caption{Comparison of the spatial distribution of injected pairs on the equatorial plane for different degrees of anisotropy, corresponding to the equatorial-plane of Fig.~\ref{Anisotropy1}.}
\label{Anisotropy2}
\end{figure*}

Fig.~\ref{GR2} shows the spatial distribution of injected plasma on the equatorial plane for the Minkowski, Schwarzschild, and Kerr metric ($a = 0.5$ and $0.9375$). Fig.~\ref{Anisotropy2} shows the spatial distribution on the equatorial plane for different degrees of anisotropy ($\xi = 0.1$, $0.25$, $1.0$, and $1000$) in the Kerr metric with spin parameter $a = 0.9375$.

\section{the Validity of the Optically Thin Approximation}\label{Appendix:optical depth}
{In this paper, we assume that photons emitted from the reconnection region propagate without undergoing collisions and follow null geodesics (see Section~\ref{Model:pair production}). In this appendix, we show the validity of this assumption through an order-of-magnitude estimate of the optical depth.}

{Magnetic reconnection in the vicinity of the black hole efficiently produces non-thermal particles and a photon field, as described by Eq.~\eqref{eq:Photon spectral}. The high-energy photon population in the power-law tail, characterized by the spectral index $p_{\rm tail}$, contributes to pair loading into the jet (see Section~\ref{results for M87}). Using Eq.~\eqref{eq:Photon spectral}, the number density of these high-energy photons can be expressed as
\begin{equation}
\tilde{n}_{\gamma}\approx \frac{\tilde{L}_{\rm rec}}{4\pi l_{\rm rec}^2 c\tilde{E}_{\rm syn}}\left(\frac{\tilde{E}_\gamma}{\tilde{E}_{\rm syn}}\right)^{(1-p_{\rm tail})/2}, \label{eq: photon number density}
\end{equation}
where $\tilde{E}_{\rm syn}=he\tilde{B}\gamma_{\rm syn}^2/(2\pi m_e c)\approx16\,\tilde{\beta}_{\rm rec,-1}\,{\rm MeV}$ is the synchrotron burnoff photon energy.
For a photon of energy $\tilde{E}_1$, the photons that contribute most efficiently to pair production have energies $\tilde{E}_2$ satisfying the threshold condition $\tilde{E}_1\tilde{E}_2 \sim (m_e c^2)^2$.
Furthermore, for a power-law photon distribution, each logarithmic energy interval contributes approximately equally to the pair-production rate for photons with energies in the range ${(m_e c^2)^2}/{\tilde{E}_{\rm syn}}<\tilde{E}_1<\tilde{E}_{\rm syn}$ (see Fig.~\ref{injection spectral}). The photon population that contributes most efficiently to pair production with photons of energy $\tilde{E}_{\rm syn}$ consists of photons with energies
${(m_e c^2)^2}/{\tilde{E}_{\rm syn}}$. The number density of these target photons is given by
\begin{equation}
  \tilde{n}_{\gamma,2}\approx  \frac{\tilde{L}_{\rm rec}}{4\pi l_{\rm rec}^2 c\tilde{E}_{\rm syn}}\left(\frac{m_e c^2}{\tilde{E}_{\rm syn}}\right)^{(1-p_{\rm tail})}\approx5\times10^8\,\tilde{B}_3^2\,\tilde{\beta}^{1.7}_{\rm rec,-1}\,{\rm cm^{-3}}.\label{eq:target numberdensity}
\end{equation}
In deriving the numerical estimate, we adopt $p_{\rm tail}=2.7$. The optical depth for $\gamma\gamma$ interactions can be roughly estimated as
\begin{equation}
\tau_{\gamma\gamma}\approx \tilde{n}_{\gamma,2}\sigma_{\gamma\gamma}r_g
\approx2\times10^{-3}\,\tilde{B}_3^2\,\tilde{\beta}_{\rm rec,-1}^{1.7}\,M_9 \ll 1,
\end{equation}
where we approximate the cross section as $\sigma_{\gamma\gamma}\approx0.2\sigma_T$ \citep{Coppi1990}.
Since the optical depth is much smaller than unity, only a small fraction of photons undergo pair production, and the vast majority propagate along geodesics without interaction. This estimate considers pair production only between high-energy photons produced by magnetic reconnection. In principle, additional electromagnetic interactions may occur, such as $\gamma\gamma$ interaction with photons from the accretion flow or corona, and Compton scattering off pairs in the magnetosphere.
However, the number density of photons from the accretion flow and corona is expected to be significantly smaller than that given by Eq.~\eqref{eq:target numberdensity}, because of larger emission regions and lower luminosities (\citet{Levinson2011L,Moscribradzka11,Kimura+20}). Furthermore, as suggested in Section~\ref{sec:result}, the pair density in the magnetosphere is at most $n_{\pm}\approx\dot{n}_{\pm}(r_g/c)\sim10^5\,{\rm cm^{-3}}$. Therefore, these processes remain optically thin and are not expected to contribute significantly compared to pair production induced by photons originating from the reconnection region.}

\section{Full Dynamic Range of the Pair Multiplicity Distribution}\label{Appendix:spartial distribution}

\begin{figure*}[t]
\centering
\includegraphics[width=\textwidth]{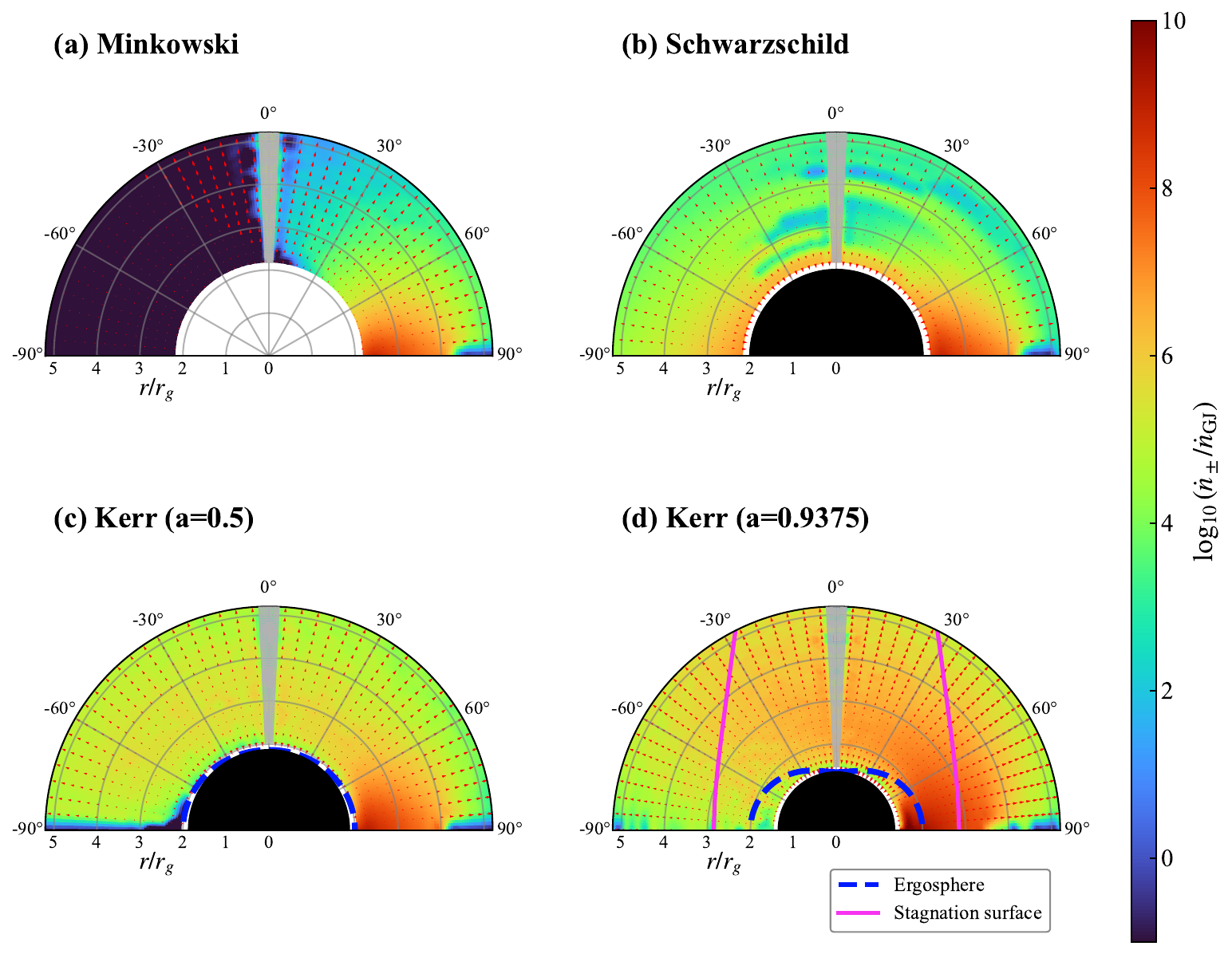}
\caption{Full dynamic-range maps of the pair production rate corresponding to Fig.~\ref{GR1}. The color scale is extended to lower multiplicities than in figure \ref{GR1}.}
\label{M87_GR_appendix}
\end{figure*}

\begin{figure*}[t]
\centering
\includegraphics[width=\textwidth]{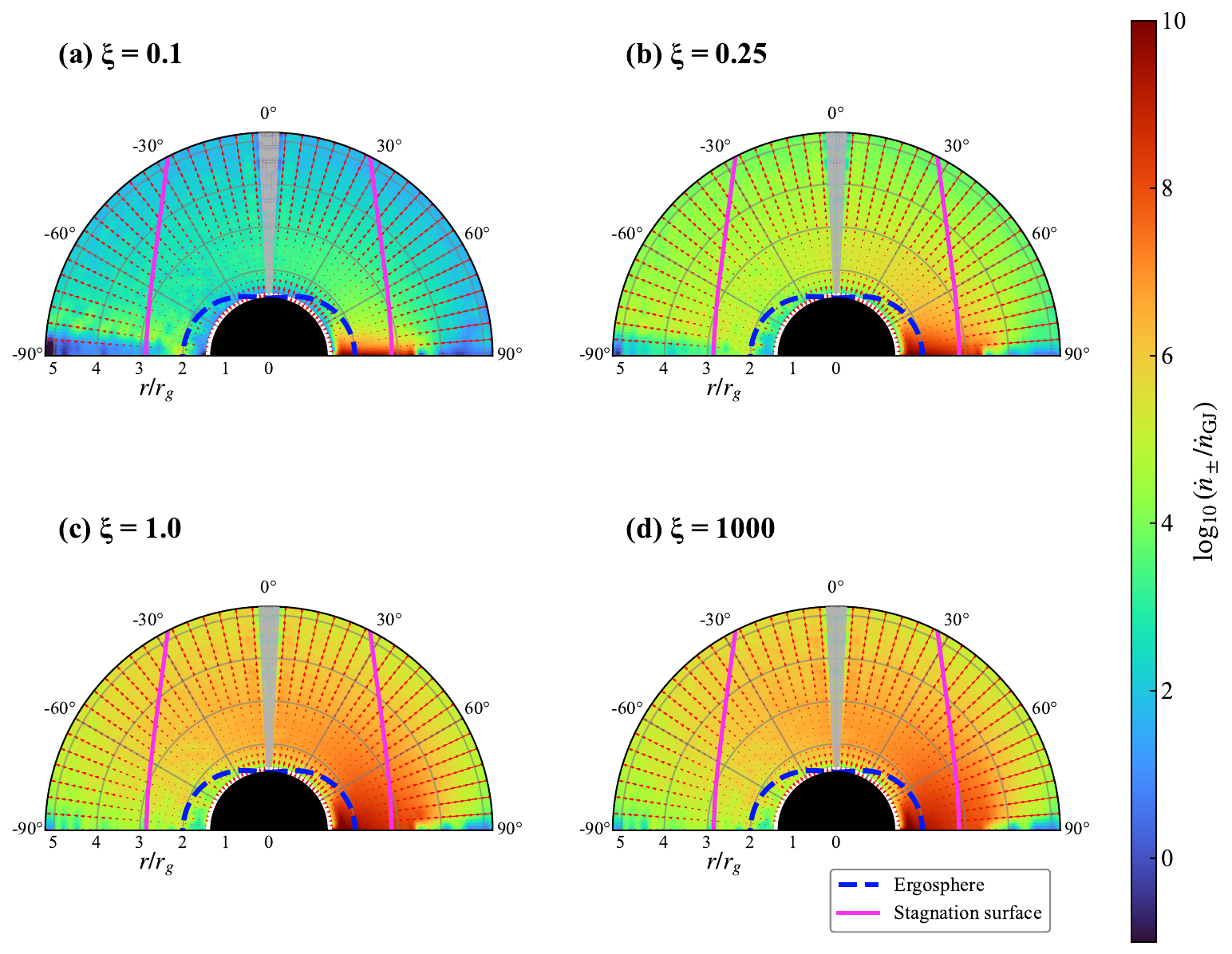}
\caption{Full dynamic-range maps of the pair production rate corresponding to Fig.~\ref{Anisotropy1}. For the adopted anisotropic parameter range, the entire jet region remains populated with injected pairs, with no completely empty regions, and the pair multiplicity is typically $\dot{n}_{\pm}/\dot{n}_{\rm GJ}\gtrsim10^{3}$--$10^{4}$ throughout the domain.}
\label{M87_kappa_appendix}
\end{figure*}

{To clarify the pair distribution below the color-scale limit adopted in Figs.~\ref{GR1} and \ref{Anisotropy1}, we present full dynamic-range maps of the pair ptroduction rate in  Figs.~\ref{M87_GR_appendix} and \ref{M87_kappa_appendix}. These figures show that, when general relativistic effects are taken into account, injected pairs populate the jet throughout the entire computational domain, leaving no completely empty regions and sustaining a pair multiplicity of at least $\dot{n}_{\pm}/\dot{n}_{\rm GJ}\gtrsim 10^{3}$--$10^{4}$. Even when anisotropic photon distributions are taken into account, the pair multiplicity remains well above unity throughout the BZ jet region. This suggests that the injected pairs can effectively screen the electric field and that our scenario is capable of closing spark gaps in the black hole magnetosphere (see Section~\ref{Discussion:VHE}).}

\bibliography{sample701}{}
\bibliographystyle{aasjournalv7}



\end{document}